\documentclass[showpacs,preprintnumbers,amsmath,amssymb,prb,twocolumn,superscriptaddress,footinbib]{revtex4}
\usepackage{graphicx}
\usepackage{bm}
\usepackage{dsfont}

\newcommand{\bk}{{\boldsymbol{k}}}
\newcommand{\bkhat}{{\boldsymbol{\hat{k}}}}
\newcommand{\bkbar}{{\boldsymbol{\bar{k}}}}

\newcommand{\bp}{{\boldsymbol{p}}}
\newcommand{\bpbar}{{\boldsymbol{\bar{p}}}}
\newcommand{\bphat}{{\boldsymbol{\hat{p}}}}
\newcommand{\cdg}{c^\dagger_{\bk \sigma}}
\newcommand{\cdgup}{c^\dagger_{\bk \uparrow}}
\newcommand{\cdgdn}{c^\dagger_{\bk \downarrow}}
\newcommand{\cu}{c_{\bk \sigma}}

\newcommand{\ex}[1]{\mathrm{e}^{#1}}
\newcommand{\im}{\mathrm{i}}

\newcommand{\du}{\mathrm{d}}

\newcommand{\cdgtd}[1]{\tilde{c}^\dagger_{#1,\bk}}
\newcommand{\ctd}[1]{\tilde{c}_{#1,\bk}}
\newcommand{\A}{\mathrm{A}}
\newcommand{\B}{\mathrm{B}}
\newcommand{\T}{\mathrm{T}}
\newcommand{\N}{\mathrm{N}}
\newcommand{\bB}[1]{{\boldsymbol{B}}_{#1}}
\newcommand{\hatbB}[1]{{\boldsymbol{\hat{B}}}_{#1}}
\newcommand{\bb}{{\boldsymbol{b}}_{\bk_\parallel}}
\newcommand{\bbp}{{\boldsymbol{b}}_{\bp_\parallel}}

\newcommand{\bsigma}{\boldsymbol{\sigma}}
\newcommand{\bo}[1]{{\boldsymbol{#1}}}
\newcommand{\unit}{\mathds{1}}
\newcommand{\Gfour}{\mathds{G}}
\newcommand{\Tr}{\mathrm{Tr}}

\begin{document}


\title{Using Josephson junctions to determine the pairing state\\ of
  superconductors without crystal inversion symmetry}


\author{K. B{\o}rkje}
\affiliation{Department of Physics, Norwegian University of Science
  and Technology, N-7491 Trondheim, Norway}


\date{\today}

\begin{abstract}
  Theoretical studies of a planar tunnel junction between two
  superconductors with antisymmetric spin-orbit coupling are
  presented. The half-space Green's function for such a superconductor
  is determined. This is then used to derive expressions for the
  dissipative current and the Josephson current of the
  junction. Numerical results are presented in the case of the Rashba
  spin-orbit coupling, relevant to the much studied compound
  CePt$_3$Si. Current-voltage diagrams, differential conductance and
  the critical Josephson current are presented for different
  crystallographic orientations and different weights of singlet and
  triplet components of the pairing state. The main conclusion is that
  Josephson junctions with different crystallographic orientations may
  provide a direct connection between unconventional pairing in
  superconductors of this kind and the absence of inversion symmetry
  in the crystal.
\end{abstract}


\maketitle



\section{Introduction}

The question of how parity violation affects superconductivity
has until recently not been subject to much experimental
studies. In
recent years, however, superconductivity has been discovered in several
materials with a noncentrosymmetric crystal structure. This offers an arena
for the study of superconductivity in the absence of inversion
symmetry. Theoretical studies of such systems have predicted several exotic
features, reviewed in Refs. \onlinecite{fujimoto_review} and
\onlinecite{sigrist_review}. The absence of inversion symmetry allows
an antisymmetric spin-orbit coupling in the Hamiltonian. This has,
among other things, the
consequence that the pairing state of the superconductor may not be
classified as a spin singlet or a spin triplet state.\cite{edelstein,gorkov}

The most famous and studied example of the noncentrosymmetric superconductors is the heavy
fermion compound CePt$_3$Si, which possess several interesting
properties.\cite{bonalde_izawa,yogi,fujimoto_review,sigrist_review}
For instance, the pairing state of CePt$_3$Si seems to contain line
nodes\cite{bonalde_izawa} even though NMR measurements are of the kind
expected for a conventional superconductor.\cite{yogi} Several theories
have been put
forward to explain this.\cite{samokhin_symm_nodes,hayashi,fujimoto_AF_nodes}
Other examples of noncentrosymmetric superconductors are UIr, Li$_2$Pd$_3$B,
Li$_2$Pt$_3$B, Cd$_2$Re$_2$O$_7$ and possibly KOs$_2$O$_6$. The absence of inversion
symmetry in these materials destroys spin degeneracy through antisymmetric spin-orbit
coupling. This is expected to be strong in some of the materials
mentioned above,\cite{samokhin,eguchi} especially in compounds
containing atoms with a large atomic number. Line nodes also seem
to appear in the pairing state of Li$_2$Pt$_3$B,\cite{yuan,nishiyama} whereas
Cd$_2$Re$_2$O$_7$,
Li$_2$Pd$_3$B and KOs$_2$O$_6$ appear to be nodeless.\cite{vyaselev_lumsden,yuan,shibauchi_shimono}

The experiments performed on these materials so far mostly concern
quantities such as specific heat, magnetic penetration depth
and the nuclear spin-lattice relaxation rate. They are all important
in order to determine the pairing state of a superconductor. However, tunneling
spectroscopy and experiments on Josephson junctions are
also a very useful tool in this respect, both in conventional and
high-$T_\text{c}$ superconductors.\cite{giaever,kashiwaya,tsuei} 

Recently, theoretical studies of transport in a junction between a normal metal and a noncentrosymmetric
superconductor were performed.\cite{yokoyama,iniotakis} These kind of transport
measurements do not probe bulk properties directly, but will depend
on how the pairing state is affected by the surface. Due to the possible triplet
component, one may expect the gap to deviate from its bulk value\cite{ambegaokar2,buchholtz2} and formation of Andreev bound states near the
surface.\cite{hu_ABS,kashiwaya}

The transport properties of a Josephson junction consisting of two
noncentrosymmetric superconductors have been
investigated in Ref. \onlinecite{borkje}. Given particular pairing states, it was noted that both
the quasiparticle current and the critical Josephson current would
depend on the relative crystal orientation of the
superconductors. Similar effects may appear with the two-band
superconductor MgB$_2$.\cite{graser,brinkman_agterberg} However, in
Ref. \onlinecite{borkje}, the bulk density of states was used, neglecting the effect of surface scattering. One might therefore
question the validity of these results, since the effect of surface
reflection was not considered.

In this paper, the effect of surface scattering is taken into account
when determining the transport properties of the above mentioned Josephson junction. It is shown that the effects
predicted in Ref. \onlinecite{borkje} may still appear, even though the
surface provides a strong coupling between the spin-orbit split
bands. Thus, in some cases one may expect qualitative changes in the
differential conductance for different relative crystal orientations of the two superconductors. In addition,
quantitative changes in the critical Josephson current may be
expected. This could make
it easier to establish a direct connection between the unconventional pairing and the absence of
inversion symmetry. This paper is hence an attempt to motivate
experimental work on such junctions. 

The paper is organized as follows. In section \ref{sec:model}, we
define the model containing a general antisymmetric spin-orbit coupling. The Green's function is then established, first in
the bulk case and then in a half-space or semi-infinite
scenario. Expressions for the tunneling currents are presented in section
\ref{sec:current_theory}. In section \ref{sec:numerical_results}, numerical
results using the Rashba spin-orbit coupling are presented to exemplify the predicted effects. 

\section{\label{sec:model} Model}

We will start by considering the bulk properties of a clean superconductor with spin-orbit
split bands. The model will be written down in the continuum
limit. Having established the bulk Green's function, we move on to
derive the Green's function in the presence of a reflecting surface.

\subsection{\label{sec:Bulk}  Bulk properties}

Let the Hamiltonian consist of two terms, $H = H_{\mathrm{N}} +
H_{\mathrm{SC}}$, a normal part and a part describing superconductivity. In the
bulk, the normal part is
\begin{equation}
  \label{eq:Hamnormal}
  H_{\mathrm{N}} = \int \du \bk \,  \phi^\dagger_{\bk}
  \left[(\varepsilon_\bk - \mu) \mathds{1} + \bB{\bk} \cdot
    \bsigma \right] \phi_\bk,
\end{equation}
where $\phi^\dagger_\bk = (\cdgup, \, \cdgdn)$, $\varepsilon_\bk$ is
the band dispersion and $\mu$ the chemical potential. The vector $\boldsymbol{\sigma}$ consists of
the three Pauli matrices. 

The vector $\bB{\bk}$ describes the
anti-symmetric spin-orbit coupling. It removes the spin
degeneracy from the band $\varepsilon_\bk$. The absence of inversion
symmetry is reflected in the property $\bB{-\bk} = - \bB{\bk}$. An electron in a state with momentum $\bk$ will align its spin parallel or
antiparallel to $\bB{\bk}$. The symmetries of $\bB{\bk}$
may be determined from point group symmetry considerations.\cite{frigeri}

Diagonalization of equation \eqref{eq:Hamnormal} gives $H_{\mathrm{N}} =
\sum_{\lambda=\pm,\bk}\xi_{\lambda,\bk} \cdgtd{\lambda}
\ctd{\lambda}$, where $\xi_{\pm,\bk} = \varepsilon_\bk - \mu \pm
|\bB{\bk}|$. The spin of an electron in a state with momentum $\bk$
will point parallel(antiparallel) to $\bB{\bk}$ in band $+$($-$).

We write down the term responsible for superconductivity in terms of the long-lived excitations in the normal
state, {\it i.e.}
\begin{equation}
  \label{eq:HamSC}
  H_{\mathrm{SC}} = \frac{1}{2} \sum_{\lambda\mu} \int \du \bk \, \du
  \bk' \, V_{\lambda \mu}(\bk,\bk') \, \tilde{c}^\dagger_{\lambda,-\bk} \tilde{c}^\dagger_{\lambda,\bk} \tilde{c}_{\mu,\bk'} \tilde{c}_{\mu,-\bk'}.
\end{equation}
We will consider
the limit where the spin-orbit splitting is much
larger than the superconducting gaps. This is a relevant limit, at
least for the materials CePt$_3$Si\cite{samokhin} and Cd$_2$Re$_2$O$_7$.\cite{eguchi} In that case, interband Cooper
pairs are strongly suppressed, even though the two bands may
touch at some isolated points on the Fermi surface.\cite{samokhin}
Thus, the model \eqref{eq:HamSC} contains only intraband Cooper pairing. However, it does include an internal Josephson coupling, {\it i.e.} scattering of Cooper pairs between the bands.

The standard mean field approach gives
\begin{equation}
  \label{eq:HamSCMF}
  H_{\mathrm{SC}} = \frac{1}{2} \sum_{\lambda} \int \du \bk \left(
    \tilde{\Delta}_{\lambda,\bk} \tilde{c}^\dagger_{\lambda,\bk}
    \tilde{c}^\dagger_{\lambda,-\bk} +  \tilde{\Delta}^\ast_{\lambda,\bk} \tilde{c}_{\lambda,-\bk}
    \tilde{c}_{\lambda,\bk} \right) 
\end{equation}
where $\tilde{\Delta}_{\lambda,\bk} = - \sum_{\mu} \int \du \bk' V_{\lambda \mu} (\bk, \bk') \langle
\tilde{c}_{\mu,\bk'} \tilde{c}_{\mu,-\bk'}
\rangle$. $\tilde{\Delta}_{\lambda,-\bk} = - \tilde{\Delta}_{\lambda,\bk}$ follows
from the fermionic anticommutation relations. One should note that the
two bands are decoupled in the mean field approximation. However, the gaps
$\tilde{\Delta}_{\pm,\bk}$ are in general not independent, but related through the self-consistency equations due to
the above mentioned possibility of interband pair scattering.\cite{mineev}

Let ${\cal K}$ denote the time-reversal operator, whose effect on the
operators in the spin basis is ${\cal
  K}: c^\dagger_{\bk,\sigma} = -\sigma c^\dagger_{-\bk,-\sigma}$. It
may be derived that ${\cal K}: \tilde{c}^\dagger_{\lambda,\bk} = t_{\lambda,\bk}
\tilde{c}^\dagger_{\lambda,-\bk}$, where $t_{\lambda,\bk} = - t_{\lambda,-\bk}$ is a
gauge-dependent phase factor. One may write $\tilde{\Delta}_{\lambda,\bk} =
t_{\lambda,\bk} \chi_{\lambda,\bk}$, where $\chi_{\lambda,\bk}$ is the
order parameter for pairs of time reversed states on which observable
quantities will depend. Thus,
$\chi_{\lambda,\bk} = \chi_{\lambda,-\bk}$ may be expanded in terms of
even basis functions of irreducible representations of the space group.\cite{sergienko}

Define the matrix $\Delta_\bk$ whose elements are the gap functions
$\Delta_{\bk,\sigma\sigma'}$ in a spin basis. By transforming equation
\eqref{eq:HamSCMF}, one arrives at
\begin{equation}
  \label{eq:gapOrig}
  \Delta_\bk =  \eta_{\bk,\mathrm{S}} \, (-\im \sigma_y) + \eta_{\bk,\mathrm{T}} ( \hatbB{\bk} \cdot \bsigma ) \, (-\im \sigma_y) .
\end{equation}
Thus, in the absence of spatial inversion
symmetry, the order parameter in a spin basis have no definite
parity, but is in general a linear combination of a singlet (S) and
a triplet (T) part. \cite{edelstein,gorkov,fujimoto,hayashi} The singlet
and triplet components are determined by 
\begin{align}
 \label{eq:singlet_triplet}
 \eta_{\bk,\mathrm{S}} & = \frac{1}{2}\left(\chi_{+,\bk} +
   \chi_{-,\bk} \right), \nonumber \\
 \eta_{\bk,\mathrm{T}} & = \frac{1}{2}\left(\chi_{+,\bk} - \chi_{-,\bk} \right).
\end{align}
There is no need to specify the momentum
dependence of the gaps $\chi_{\lambda,\bk}$ at this point. 

In the bulk, the Green's functions are diagonal in momentum space due
to translational symmetry. In the imaginary time formalism, define the normal and anomalous Green's
functions as ${\cal G}_{\text{b}, \sigma \sigma'}(\bk,\tau) = - \langle
T_\tau \, c_{\bk
  \sigma}(\tau) \, c^\dagger_{\bk \sigma'} (0) \rangle$ and ${\cal F}_{\text{b}, \sigma
  \sigma'}(\bk,\tau) = \langle T_\tau \, c_{\bk \sigma}(\tau) \, c_{-\bk
  \sigma'} (0) \rangle$, respectively, where the subscript b denotes
bulk. It is convenient to transform to fermionic Matsubara
frequencies, $\omega_n = (2n+1)\pi/\beta$, where $\beta$ is the
inverse temperature. 
The bulk Green's function in spin $\times$ particle-hole space,
\begin{equation}
  \label{eq:full_bulk}
  \Gfour_\text{b} (\bk,\im \omega_n) = \left( \begin{array}{cc}
   {\cal G}_\text{b} (\bk,\im \omega_n) &  - {\cal F}_\text{b} (\bk, \im \omega_n) \vspace{0.2cm} \\ 
   - {\cal F}^\dagger_\text{b} (\bk, \im \omega_n) &  -{\cal G}^\text{t}_\text{b}
   (-\bk,-\im \omega_n) \end{array} \right),
\end{equation}
is found by solving the Gor'kov equations \eqref{eq:Gorkov_bulk}
presented in appendix
\ref{sec:app_Green}. The components are matrices in spin space, given by
\begin{align}
  \label{eq:bulk_Green}
  {\cal G}_\text{b} (\bk, \im \omega_n) & = \frac{1}{2} \sum_{\lambda=\pm} \sigma^{\lambda}_{\hatbB{\bk}}
    G_\lambda(\bk,\im \omega_n) \, , \nonumber \\
   {\cal F}_\text{b} (\bk, \im \omega_n) & = - \frac{\im}{2}
   \sum_{\lambda=\pm} \sigma^{\lambda}_{\hatbB{\bk}} \sigma_y \, F_\lambda(\bk,\im \omega_n)  \, ,
\end{align}
in terms of the complex scalar functions
\begin{align}
 \label{eq:G,F}
 G_\lambda (\bk,\im \omega_n) & = - \frac{\im \omega_n +
   \xi_{\lambda,\bk}}{\omega_n^2 + \xi_{\lambda,\bk}^2 +
   |\chi_{\lambda,\bk}|^2} \, , \nonumber \\
 F_\lambda (\bk,\im \omega_n) & =
 \frac{\chi_{\lambda,\bk}}{\omega_n^2 + \xi_{\lambda,\bk}^2 +
   |\chi_{\lambda,\bk}|^2} \, ,
\end{align} 
and the matrices
\begin{equation}
  \label{eq:sigma_lambda_def}
  \sigma^{\lambda}_{\hatbB{\bk}} = \mathds{1} + \lambda \hatbB{\bk} \cdot \bsigma.
\end{equation}

\subsection{\label{sec:halfspace}Half-space Green's function}

The bands $+$ and $-$ defined in the previous section has the property
that reversing the direction of an electron's momentum while
preserving its spin requires a change of bands. Thus, one would expect
that the independence of the bands $+$ and $-$ could be vulnerable to
scattering, {\it e.g.} from impurities. In fact, it has been shown that a small concentration of nonmagnetic impurities
does not change the picture of independent bands in the mean field
approximation.\cite{frigeri2} A perfectly reflecting surface should
however lead to a severe mixing of the bands. This needs to
be taken into account when describing transport in heterostructures
containing these materials. 

The presence of a surface will make the Hamiltonian and the Green's function nondiagonal in
momentum space. Still, due to the nature of the spin-orbit coupling,
it is convenient to work in a plane wave basis. We will assume that
the surface is perfectly smooth. Of course, any real surface will have some roughness,
which may very well modify the results of this
paper. However, at least for not too rough surfaces, this model is an
appropriate starting point. We will also assume that the surface is
spin inactive, {\it i.e} nonmagnetic.

Consider the simplest case of a perfectly smooth surface at $x =
0$, such that the electrons are confined to $x < 0$. We seek the Green's function $\Gfour (\bk_1,\bk_2,\tau)$, whose
elements are
\begin{align}
  \label{eq:def_planewave_Green}
  {\cal G}_{\sigma \sigma'} (\bk_1,\bk_2,\tau) & \equiv
- \langle T_\tau c_{\bk_1, \sigma}(\tau) c^\dagger_{\bk_2, \sigma'}(0)
\rangle \, , \nonumber \\ {\cal F}_{\sigma \sigma'} (\bk_1,\bk_2,\tau)
& \equiv
\langle T_\tau c_{\bk_1, \sigma}(\tau) c_{-\bk_2, \sigma'}(0)\rangle \,,
\end{align}
where $c_{\bk_1, \sigma}$ is the annihilation operator for
a {\it plane wave} state. In the presence of a scattering surface, these correlation functions will not be
diagonal in momentum space. 

Due to translational invariance in the $y$- and $z$-direction, it is natural
to introduce the 4x4 Green's function in spin $\times$ particle-hole
space in a mixed representation, $\tilde{\Gfour}
(x_1,x_2,\bk_\parallel,\im \omega_n)$. We have defined $\bk_\parallel = k_y \hat{\bo{y}} + k_z
\hat{\bo{z}}$. The normal and
anomalous components are ${\cal \tilde{G}}_{\sigma \sigma'}(x_1,x_2,\bk_\parallel,\tau) = - \langle
T_\tau \, c_{x_1,\bk_\parallel, \sigma}(\tau) \, c^\dagger_{x_2,\bk_\parallel, \sigma'} (0) \rangle$ and ${\cal \tilde{F}}_{\sigma
  \sigma'}(x_1,x_2,\bk_\parallel,\tau) = \langle T_\tau \, c_{x_1,\bk_\parallel,
  \sigma}(\tau) \, c_{x_2,-\bk_\parallel, \sigma'} (0) \rangle$, respectively. The
Green's function is determined by the Gor'kov equations \eqref{eq:Gorkov1}, which are presented
in appendix \ref{sec:app_Green}. The boundary conditions are
\begin{equation}
  \label{eq:bound_bond}
  \tilde{\Gfour} (x_1,x_2,\bk_\parallel,\im \omega_n) = 0 \, , \ x_1 = 0\text{ or }x_2 = 0.
\end{equation}
The pair potential in this mixed basis, $\Delta (x_1,x_2,\bk_\parallel)$, should be determined
self-consistently. Even though it may deviate significantly from
the bulk near surfaces,\cite{ambegaokar2,buchholtz2}
we will approximate it by its bulk value. This
approximation is expected to give qualitatively correct results.\cite{kashiwaya,iniotakis} In appendix \ref{sec:app_Green} it is shown
that this approximation enables us to express the
half-space Green's function in terms of bulk Green's
functions. This may be realized by treating the surface as a wall of
nonmagnetic impurities of infinite strength.\cite{bobkov} The momentum space
Green's function then becomes \begin{widetext}
\begin{align}
  \label{eq:Green_text}
\Gfour (\bk_1,\bk_2,\im \omega_n) = \Big[ \Gfour_\text{b} (\bk_1,\im \omega_n) \delta(k_{1,x} - k_{2,x})  - \Gfour_\text{b} (\bk_1,\im \omega_n) \tilde{\Gfour}^{-1}_\text{b}
  (0,0,\bk_\parallel,\im \omega_n)  \Gfour_\text{b}
(\bk_2,\im \omega_n) \Big] \delta(\bk_{1,\parallel} - \bk_{2,\parallel}).
\end{align}
\end{widetext}
To determine this, we need the inverse of the matrix 
\begin{equation}
  \label{eq:invert_matrix}
  \tilde{\Gfour}_\text{b}
  (0,0,\bk_\parallel,\im \omega_n) = \int_{-\infty}^\infty \du k_x \Gfour_\text{b}
(\bk,\im \omega_n).
\end{equation}

Let us now define $\bk \equiv (k_x,\bk_\parallel)$ and $\bkbar \equiv
(-k_x,\bk_\parallel)$. From the previous subsection, we saw that the
gaps $\chi_{\pm,\bk}$ where unchanged upon reversal of the momentum.
At this point we restrict ourselves to surfaces such that the gaps are
unchanged also when
reversing the component of the momentum perpendicular to the surface only,
{\it i.e.} $\chi_{\lambda,\bkbar} = \chi_{\lambda,\bk}$. Although this is not a
necessary requirement to determine the Green's function, it will simplify the
calculations and be sufficient for the scenarios considered here. We will also
assume $\xi_{\lambda,\bkbar} = \xi_{\lambda,\bk}$. Using these approximations, the properties $G_\lambda (\bkbar,\im \omega_n) = G_\lambda (\bk,\im
\omega_n)$ and $F_\lambda (\bkbar,\im \omega_n) = F_\lambda
(\bk,\im \omega_n)$ follow from equations
\eqref{eq:G,F}. We now convert the $k_x$-integral in
\eqref{eq:invert_matrix} to an energy integral. The integrand
will be strongly peaked about the Fermi level. Thus, we apply the
quasiclassical approximation of replacing all momentum-dependent
quantities by their value at the Fermi level.\cite{buchholtz} We
introduce the notation 
\begin{equation}
  \label{eq:kF_Def}
  \bk_\text{F} \equiv (k_{\text{F},x},\bk_\parallel) \, , \quad \bkbar_\text{F}
  \equiv (-k_{\text{F},x},\bk_\parallel) \, ,
\end{equation}
where
$k_{\text{F},x} \geq 0$ is determined by $\xi_{\bk_\text{F}} = 0$ given $\bk_\parallel$. We define the
quasiclassical or $\xi$-integrated Green's functions by
\begin{align}
  \label{eq:quasi}
  g_{\lambda} (\bk_\text{F},\im \omega_n) & = -\frac{\im \omega_n}{\sqrt{\omega_n^2 + |\chi_{\lambda,\bk_\text{F}}|^2}} \, ,
\nonumber \\
f_{\lambda} (\bk_\text{F},\im \omega_n) & =
\frac{\chi_{\lambda,\bk_\text{F}}}{\sqrt{\omega_n^2 + |\chi_{\lambda,\bk_\text{F}}|^2}} \, .
\end{align}
The integral over the normal Green's function in matrix
\eqref{eq:invert_matrix} is found using 
\begin{equation}
  \label{eq:first_element_inv_matrix}
   \frac{1}{2} \int_{-\infty}^\infty \du k_x \, 
   \sigma^{\lambda}_{\hatbB{\bk}} G_\lambda (\bk,\im \omega_n) = \pi N^x_{\lambda,\bk_\text{F}} \sigma^{\lambda}_{\bb}
   g_{\lambda} (\bk_\text{F},\im \omega_n) ,
\end{equation}
where $N^x_{\lambda,\bk_\text{F}}$ is $|\partial_{k_x}
\xi_{\lambda,\bk}|^{-1}$ taken at $\bk_\text{F}$. The vector 
\begin{equation}
  \label{eq:bb_def}
  \bb = \frac{1}{2} (\hatbB{\bk_\text{F}} + \hatbB{\bkbar_\text{F}})
\end{equation}
has the property
${\boldsymbol{b}}_{-\bk_\parallel} = - \bb$, but is not a unit vector. Similarly, the integral
over the anomalous Green's function is obtained from
\begin{equation}
  \label{eq:second_element_inv_matrix}
   \frac{1}{2} \int_{-\infty}^\infty \du k_x \, 
   \sigma^{\lambda}_{\hatbB{\bk}}  F_\lambda (\bk,\im \omega_n) = \pi N^x_{\lambda,\bk_\text{F}} \sigma^{\lambda}_{\bb}
   f_{\lambda} (\bk_\text{F},\im \omega_n) .
\end{equation}
We now assume that the difference
in the density of states of the two spin-orbit split bands is small
and may be neglected. Consequently, we also let $N^x_{+,\bk_\text{F}} =
N^x_{-,\bk_\text{F}} \equiv N^x_{\bk_\text{F}}$.\footnote{One might
  expect that this approximation breaks down when $\bk_{\text{F},x}
  \rightarrow 0$ and that
  $N^x_{\pm,\bk_\text{F}} \rightarrow 0$ in that case, but this will
  not pose any problem here.} This is
not a necessary step in order to proceed, but it simplifies the calculations.

The inverse of the matrix \eqref{eq:invert_matrix} is then \begin{widetext}
\begin{equation}
  \label{eq:invertert_matrise}
  \tilde{\Gfour}^{-1}_\text{b}
  (0,0,\bk_\parallel,\im \omega_n) = \frac{1}{\pi K(\bk_\text{F},\im
    \omega_n) \, N^x_{\bk_\text{F}} } \sum_{\rho = \pm} \left( \begin{array}{cc}
   \sigma^{\rho}_{\bb} \, g^\ast_\rho(\bk_\text{F},\im \omega_n) &
   \im \sigma^{\rho}_{\bb} \sigma_y \, f_\rho(\bk_\text{F},\im \omega_n) \vspace{0.2cm} \\ 
   -\im \sigma_y \sigma^{\rho}_{\bb} \, f^\ast_\rho(\bk_\text{F},\im
   \omega_n) &  - \sigma_y \sigma^{\rho}_{\bb} \sigma_y \, g_\rho(\bk_\text{F},\im \omega_n) \end{array} \right).
\end{equation}
We have introduced the function
\begin{equation}
  \label{eq:Kdef}
   K(\bk_\text{F},\im \omega_n) = 2 \left[ b_{+,\bk_\parallel} +
     b_{-,\bk_\parallel} \frac{\omega_n^2 +
       \mathrm{Re}(\chi_{+,\bk_\text{F}}
       \chi^\ast_{-,\bk_\text{F}})}{\sqrt{\omega_n^2 +
         |\chi_{+,\bk_\text{F}}|^2}\sqrt{\omega_n^2 +
         |\chi_{-,\bk_\text{F}}|^2}} \right],
\end{equation}
where $b_{\pm,\bk_\parallel} = 1 \pm |\bb|^2$. Later, it will be
apparent that zeros in $K(\bk_\text{F},\im \omega_n)$ will correspond
to surface bound states.

Introduce the simplified notation
$G_{\lambda,1} \equiv G_\lambda(\bk_1,\im \omega_n)$,
$F_{\mu,2} \equiv F_\mu(\bk_2,\im \omega_n)$ and $g_{\rho} \equiv g_\rho
(\bk_\text{F},\im \omega_n)$. No momentum index is needed on the latter since
it depends only on the parallel momentum and $\bk_{1,\parallel} = \bk_{2,\parallel} \equiv \bk_\parallel$. We are then ready to write down the half-space Green's
function. The normal and anomalous parts, defined in
\eqref{eq:def_planewave_Green}, are
\begin{align}
  \label{eq:G_half_space}
  {\cal G}(\bk_1,\bk_2,\im \omega_n) & = \frac{1}{2} \Bigg\{ \sum_{\lambda} \sigma^{\lambda}_{\hatbB{\bk}} G_{\lambda,1}
  \delta(k_{1,x} - k_{2,x}) \\ 
& - \frac{1}{2 \pi K(\bk_\text{F},\im \omega_n)
    N^x_{\bk_\text{F}}}
  \sum_{\lambda \rho \mu} \tilde{\sigma}^{\lambda \rho
    \mu}_{\bk_1,\bk_2} \Big[G_{\lambda,1}
    (g^{\ast}_{\rho} G_{\mu,2} + f_{\rho} F^{\ast}_{\mu,2}) +
    F_{\lambda,1} (f^{\ast}_{\rho} G_{\mu,2} - g_{\rho}
    F^{\ast}_{\mu,2} ) \Big] \Bigg\} \delta(\bk_{1,\parallel} -
  \bk_{2,\parallel}) \nonumber 
\end{align}
and
\begin{align}
  \label{eq:F_half_space}
  {\cal F}(\bk_1,\bk_2,\im \omega_n) & = - \frac{\im}{2} \Bigg\{
  \sum_{\lambda} \sigma^{\lambda}_{\hatbB{\bk}} \sigma_y \, F_{\lambda,1}
  \delta(k_{1,x} - k_{2,x}) \\ 
& - \frac{1}{2 \pi K(\bk_\text{F},\im \omega_n)
    N^x_{\bk_\text{F}}}
  \sum_{\lambda \rho \mu} \tilde{\sigma}^{\lambda \rho
    \mu}_{\bk_1,\bk_2} \sigma_y \, \Big[G_{\lambda,1}
    (g^{\ast}_{\rho} F_{\mu,2} - f_{\rho} G^{\ast}_{\mu,2}) + F_{\lambda,1}
    (f^{\ast}_{\rho} F_{\mu,2} + g_{\rho}  G^{\ast}_{\mu,2}) \Big]
  \Bigg\} \delta(\bk_{1,\parallel} - \bk_{2,\parallel}) \, , \nonumber 
\end{align}\end{widetext}
respectively. These functions are found by
inserting equations \eqref{eq:full_bulk} and
\eqref{eq:invertert_matrise} into equation \eqref{eq:Green_text}.

We have defined the matrix
\begin{equation}
  \label{eq:tildesigma_def}
  \tilde{\sigma}^{\lambda \rho
    \mu}_{\bk_1,\bk_2} = \sigma^\lambda_{\hatbB{\bk_1}}
  \sigma^\rho_{\bb} \sigma^\mu_{\hatbB{\bk_2}} \equiv \beta^{\lambda \rho
    \mu}_{\bk_1,\bk_2} \unit + \bo{\alpha}^{\lambda \rho
    \mu}_{\bk_1,\bk_2} \cdot \bo{\sigma} \, ,
\end{equation}
where the expressions for the scalar $\beta^{\lambda \rho \mu}_{\bk_1,\bk_2}$ and
the vector $\bo{\alpha}^{\lambda \rho \mu}_{\bk_1,\bk_2}$ are given in
appendix \ref{sec:app_Pauli}.

\section{\label{sec:current_theory} Calculation of tunneling currents}

Let us now consider a planar tunnel junction between two superconductors with
spin-orbit split bands. We name the systems A and B and let the
$x$-axis point perpendicular to the junction. In addition,
we use the letter $c$ for operators and $\bk$ for momenta on side A,
and $d$ and $\bp$ for the corresponding quantities on side B. The
spin-orbit coupling is described by the vectors $\bB{\bk}^\A$ and
$\bB{\bp}^\B$ on each side. These vectors are not necessarily
equal. Let us briefly
exemplify this by considering the Rashba spin-orbit coupling,
$\bB{\bk} = \alpha (\hat{\bo{n}} \times \bk)$,
even though we will work with a general $\bB{\bk}$. Here, the
vector $\hat{\bo{n}}$ describes the direction of broken inversion
symmetry of the crystal. This means that if the crystallographic orientation on side
B is different from side A, $\bB{\bk}^\A$ and $\bB{\bp}^\B$ will point
in different directions even when $\bk = \bp$.

The tunneling process is described by 
\begin{equation}
  \label{eq:tunn_Ham}
  H_\T = \sum_{\sigma \sigma'} \int \du \bk \du \bp \left({\cal T}_{\bk \bp, \sigma
      \sigma'} \cdg d_{\bp \sigma'} + {\cal T}^\ast_{\bk \bp, \sigma
      \sigma'} d^\dagger_{\bp \sigma'} \cu \right).
\end{equation}
The validity of results using perturbation theory in the tunneling Hamiltonian
formalism has been showed by Prange.\cite{prange}

We emphasize that the systems are described in terms of {\it plane wave}
states. Thus, ${\cal T}_{\bk \bp, \sigma \sigma'}$ is the transfer amplitude from
an incoming plane wave state with momentum $\bp$ on side B to an outgoing
plane wave state with momentum $\bk$ on side A. When scattering a
plane wave on a barrier, the perpendicular momentum of the transmitted
wave points in the same direction as the incoming wave. In
addition, we assume that the tunneling process conserves spin. These
properties result in 
\begin{equation}
  \label{eq:T_kp_egenskaper}
  {\cal T}_{\bk \bp, \sigma \sigma'} \equiv T_{\bk,\bp} \, \Theta[k_x p_x]
  \delta_{\sigma,\sigma'} \, ,
\end{equation}
where $\Theta[x]$ is the Heaviside step function.
Time-reversal symmetry also demands $T^\ast_{-\bk,-\bp} = T_{\bk,\bp}$.

Of course, there is also an amplitude for the incoming plane wave being reflected. However, when treating equation
\eqref{eq:tunn_Ham} as a perturbation, the current will be expressed
as Green's functions of the unperturbed systems A and B. Thus, the
reflection is taken into account by using the half-space Green's
functions obtained in the previous section.

The current from side B to side A is defined as $I(t) = -e \langle
\dot{{\cal N}}_\A \rangle$, where ${\cal N}_\A$ is the total charge
operator on side A and the operator $\dot{{\cal N}}_\A$ is given by
the Heisenberg equation $\dot{{\cal N}}_\A = \im \left[H_\T,{\cal N}_\A\right]$.
Treating the tunneling Hamiltonian as a perturbation, the Kubo formula
gives $I(t) = I_\text{qp} + I_\text{J} (t)$,\cite{mahan} where
\begin{align}
  \label{eq:current_def}
  I_\text{qp} & = -2e \, \mathrm{Im} \, \Phi(eV) \, , \nonumber \\
  I_\text{J}(t) & = 2e \, \mathrm{Im} \, \left[\ex{-2 \im eV t} \Psi(eV)
  \right] \, .
\end{align}
In the imaginary time formalism, when defining $M(\tau) \equiv
\sum_{\sigma \sigma'} \int \du \bk \du \bp \, \cdg(\tau) d_{\bp
  \sigma'}(\tau)$, we have
\begin{align}
  \label{eq:PhiPsi}
  \Phi(\im \omega_\nu) & = - \int_0^\beta \du \tau \, \ex{\im \omega_\nu
    \tau} \langle T_\tau M(\tau) M^\dagger(0) \rangle \, , \nonumber \\
  \Psi(\im \omega_\nu) & = - \int_0^\beta \du \tau \, \ex{\im \omega_\nu
    \tau} \langle T_\tau M(\tau) M(0) \rangle \, .
\end{align}
The time dependence of the operators are given by the unperturbed
Hamiltonian and the expectation values are to be taken in the
unperturbed state. The voltage is defined by $eV = \mu_\A -
\mu_\B$. The bosonic Matsubara frequency is $\omega_\nu = 2 \nu
\pi/\beta$, which will be subject to $\im \omega_\nu \rightarrow eV +
\im 0^+$. 

Equations \eqref{eq:PhiPsi} may be written as
\begin{align}
  \Phi(\im \omega_\nu) & = \frac{1}{\beta} \int \du \bk_1 \du \bk_2
  \du \bp_1 \du \bp_2 \, T_{\bk_1,\bp_1}
  T^\ast_{\bk_2,\bp_2} \label{eq:Phi_calc} \\
& \times \sum_{\omega_n} \mathrm{Tr} \Big[{\cal
      G}_\A(\bk_2,\bk_1,\im \omega_n - \im \omega_\nu) {\cal
      G}_\B(\bp_1,\bp_2,\im \omega_n) \Big] \,  \nonumber 
\end{align}
and
\begin{align}
  \Psi(\im \omega_\nu) & = \frac{1}{\beta} \int \du \bk_1 \du \bk_2
  \du \bp_1 \du \bp_2 \,
  T_{\bk_1,\bp_1} T^\ast_{\bk_2,\bp_2}  \label{eq:Psi_calc} \\
& \times \sum_{\omega_n} \mathrm{Tr} \left[{\cal
      F}^\dagger_\A(\bk_2,\bk_1,\im \omega_n - \im \omega_\nu) {\cal
      F}_\B(\bp_1,\bp_2,\im \omega_n) \right] \, ,\nonumber
\end{align}
where the components of the Green's functions are defined in
\eqref{eq:def_planewave_Green} and Tr denotes a trace in spin space.

As before, when converting the momentum integrals to energy integrals,
we replace all momentum-dependent quantities by their value on the
Fermi level. The density of states $N(\bk_\text{F})$ at the Fermi level
is assumed equal in both bands. In addition, we assume
$N(\bkbar_\text{F}) = N(\bk_\text{F})$ and
$T_{\bkbar_\text{F},\bpbar_\text{F}} = T_{\bk_\text{F},\bp_\text{F}}
$. 

\subsection{\label{sec:QP} The quasiparticle current}

Inserting the Green's function \eqref{eq:G_half_space} in equation \eqref{eq:Phi_calc}, one arrives at 
\begin{align}
  \label{eq:Phi_total}
  \Phi(\im \omega_\nu) & = \frac{\pi^2}{4} \int^\prime \du \bkhat \int^\prime \du
  \bphat \, |T_{\bk_\text{F},\bp_\text{F}}|^2
  N^\A(\bk_\text{F}) N^\B(\bp_\text{F}) \nonumber \\
  & \times \Bigg\{A_1^{\lambda \gamma} (\bk_\text{F},\bp_\text{F})
    S_1^{\lambda \gamma} (\bk_\text{F},\bp_\text{F},\im \omega_\nu)
     \\
 & \quad - \frac{1}{2} A_2^{\lambda \gamma \eta \nu}(\bk_\text{F},\bp_\text{F})
   S_2^{\lambda \gamma \eta \nu}
  (\bk_\text{F},\bp_\text{F},\im \omega_\nu)  \nonumber \\
 & \quad - \frac{1}{2} A_3^{\lambda \rho \mu \gamma}
  (\bk_\text{F},\bp_\text{F}) S_3^{\lambda \rho \mu \gamma}
  (\bk_\text{F},\bp_\text{F},\im \omega_\nu)  \nonumber \\
 & \quad + \frac{1}{4} A_4^{\lambda \rho \mu \gamma \eta \nu} (\bk_\text{F},\bp_\text{F})
  S_4^{\lambda \rho \mu \gamma \eta \nu}
  (\bk_\text{F},\bp_\text{F},\im \omega_\nu) \Bigg\} \, , \nonumber 
\end{align}
where {\it repeated greek indices are to be summed over}. The prime indicates that
the integrals over the Fermi surfaces are restricted to positive
$\hat{k}_x,\hat{p}_x$. The $A_i$'s are defined by\footnote{There
  should also be an index A(B) on the $\tilde{\sigma}$'s, since they
  depend on $\hatbB{\bk(\bp)}^{\A(\B)}$. We skip
  this, since the momenta ($\bk$ or $\bp$) tells us to which side they belong.} 
\begin{align}
\label{eq:A_def}
   & A_1^{\lambda \gamma}(\bk,\bp) = 
    \mathrm{Tr} \, \sigma^{\lambda}_{\hatbB{\bk}^\A}
    \sigma^\gamma_{\hatbB{\bp}^\B}  + \mathrm{Tr} \,  \sigma^{\lambda}_{\hatbB{\bkbar}^\A}
    \sigma^\gamma_{\hatbB{\bpbar}^\B}  \, , \nonumber \\
  & A_2^{\lambda \gamma \eta \nu}
  (\bk,\bp) =  \Tr \,
    \sigma^{\lambda}_{\hatbB{\bk}^\A} \tilde{\sigma}^{\gamma \eta
      \nu}_{\bp,\bp} + \Tr \, \sigma^{\lambda}_{\hatbB{\bkbar}^\A} \tilde{\sigma}^{\gamma \eta
      \nu}_{\bpbar,\bpbar}  \, , \nonumber \\
  & A_3^{\lambda \rho \mu \gamma} (\bk,\bp) = 
    \Tr \, \tilde{\sigma}^{\lambda \rho \mu}_{\bk,\bk}
    \sigma^{\gamma}_{\hatbB{\bp}^\B} +\Tr \, \tilde{\sigma}^{\lambda \rho \mu}_{\bkbar,\bkbar}
    \sigma^{\gamma}_{\hatbB{\bpbar}^\B} \, , \\
  & A_4^{\lambda \rho \mu \gamma \eta \nu} (\bk,\bp) =
    \Tr \, \tilde{\sigma}^{\lambda \rho \mu}_{\bk,\bk} \tilde{\sigma}^{\gamma \eta
      \nu}_{\bp,\bp} + \Tr \, \tilde{\sigma}^{\lambda \rho \mu}_{\bk,\bkbar} \tilde{\sigma}^{\gamma \eta
      \nu}_{\bpbar,\bp} \nonumber  \\
 & \qquad \qquad \qquad   + \Tr \, \tilde{\sigma}^{\lambda \rho \mu}_{\bkbar,\bk} \tilde{\sigma}^{\gamma \eta
      \nu}_{\bp,\bpbar} + \Tr \, \tilde{\sigma}^{\lambda \rho \mu}_{\bkbar,\bkbar} \tilde{\sigma}^{\gamma \eta
      \nu}_{\bpbar,\bpbar} . \nonumber 
\end{align}
These quantities depend on
$\hatbB{\bk}^\A,\hatbB{\bkbar}^\A,\hatbB{\bp}^\B,\hatbB{\bpbar}^\B$
and explicit expressions are given in appendix \ref{sec:app_Pauli}. The
$S_i$'s depend on the momenta through the gaps and are defined as
\begin{align}
\label{eq:S_def}
  & S_1^{\lambda \gamma}(\bk,\bp,\im \omega_\nu) = \frac{1}{\beta} \sum_{\omega_n}
  g^\A_\lambda(\bk,\im \omega_n - \im \omega_\nu) g^\B_\gamma(\bp,\im
  \omega_n) \, , \nonumber \\
   & S_2^{\lambda \gamma \eta \nu}
  (\bk,\bp,\im \omega_\nu)= \frac{1}{\beta} \sum_{\omega_n}
  g^\A_\lambda(\bk,\im \omega_n - \im \omega_\nu) \Gamma^\B_{\gamma \eta
    \nu} (\bp,\im \omega_n) \, , \nonumber \\
  & S_3^{\lambda \rho \mu \gamma} (\bk,\bp,\im \omega_\nu) =
  \frac{1}{\beta} \sum_{\omega_n} \Gamma^\A_{\lambda \rho \mu}(\bk,\im
  \omega_n - \im \omega_\nu) g^\B_\gamma(\bp,\im \omega_n) \, , \nonumber \\
  & S_4^{\lambda \rho \mu \gamma \eta \nu}
  (\bk,\bp,\im \omega_\nu) \\
 & \qquad \qquad = \frac{1}{\beta} \sum_{\omega_n} \Gamma^\A_{\lambda \rho \mu}(\bk,\im
  \omega_n - \im \omega_\nu) \Gamma^\B_{\gamma \eta
    \nu} (\bp,\im \omega_n). \nonumber 
\end{align}
The function $g_\lambda(\bk,\im \omega_n)$ was defined in equation
\eqref{eq:quasi}. The function $\Gamma_{\lambda \rho \mu}(\bk,\im \omega_n)$ is
\begin{align}
  \label{eq:Gamma_def}
  \Gamma_{\lambda \rho \mu}(\bk,\im \omega_n) = \frac{g_\lambda \left[g^\ast_\rho g_\mu + f_\rho
      f^\ast_\mu \right]  + f_\lambda \left[f^\ast_\rho g_\mu - g_\rho
      f^\ast_\mu \right]}{K(\bk,\im
    \omega_n)}\, ,
\end{align}
where the arguments of the $g$'s and
$f$'s were omitted for clarity. Note that $\Gamma_{\lambda \rho
  \mu}(\bk_\text{F},\im \omega_n)$ does not depend on $N^x_{\bk_\text{F}}$.

In equation \eqref{eq:Phi_total}, we have reached the point at which
the current $I_\text{qp}$ is expressed as two surface integrals over half
of the Fermi surface on each side. In addition, one is left with the
Matsubara sums which may be converted to energy integrals. To get further, one must insert the appropriate
angular dependence of the quantities $\chi_{\lambda,\bk_\text{F}}$,
$\hatbB{\bk_\text{F}}$, $N(\bk_\text{F})$ on each side as well as $|T_{\bk_\text{F},\bp_\text{F}}|$. In most cases, the remaining
integrals need to be performed numerically. Both the energy and angle
integrands contain integrable singularities which must be handled with care.

We will now assume that the two gaps $\chi_{\pm,\bk}$ are phase-locked due
to the internal Josephson coupling. We write out the phase explicitly, such that $\chi^\A_{\pm, \bk}
\rightarrow \chi^\A_{\pm,\bk} \ex{\im
  \vartheta^\A}$. $\chi^\A_{+,\bk}$ and $\chi^\A_{-,\bk}$ are real
from now on, but not necessarily of the same sign. Obviously, the same also
applies to the gaps on side B. 

To obtain the current $I_\text{qp}$, we need $\mathrm{Im} \, \Phi(\im
\omega_\nu)$. Since the $A_i$'s are real (see appendix \ref{sec:app_Pauli}), the only complex parts are
contained in the Matsubara sums.
By converting the sums to
contour integrals in the complex plane and deforming the contour, one
finds that $\mathrm{Im} \, S_i(eV + \im 0^+)$ may be expressed as energy integrals containing the functions $\mathrm{Im} \, g_\lambda(\bk,E+\im 0^+)$,
$\mathrm{Im} \, \Gamma_{\lambda \rho \mu}(\bk,E+\im 0^+)$ and the
Fermi-Dirac distribution $n_\text{F}(E)$. Details of this
procedure and the choice of appropriate branch cuts are found in
appendix \ref{sec:app_contour}. The first function is proportional to the usual bulk density of states
\begin{equation}
  \label{eq:Im_g}
  \mathrm{Im} \, g_\lambda(\bk,E+\im 0^+) = - \Theta\Big[|E| - |\chi_{\lambda,\bk}|\Big]\frac{|E|}{\sqrt{E^2 - |\chi_{\lambda,\bk}|^2}} 
\end{equation}
As before, $\Theta[x]$ is the Heaviside step function.
The second function is somewhat complicated, but may be written as\begin{widetext}
\begin{align}
  \label{eq:Im_Gamma}
  \mathrm{Im} \, \Gamma_{\lambda \rho \mu}(\bk,E+\im 0^+)  =
  \Theta\Big[|E| - |\chi_{m,\bk}|\Big] P_{\lambda \rho \mu}(E) + \Theta\Big[|\chi_{m,\bk}| - |E|
  \Big] \Theta\Big[-\chi_{+,\bk}\chi_{-,\bk}\Big] \bar{P}_{\lambda \rho
    \mu}(E) \, \delta(|E|-E_{0,\hat{\bk}})  \, ,
\end{align}
where $|\chi_{m,\bk}| \equiv
\text{min}(|\chi_{+,\bk}|,|\chi_{-,\bk}|)$. The functions $P_{\lambda \rho \mu}(E)$ and
$\bar{P}_{\lambda \rho \mu}(E)$ are even functions of $E$ and may be found
by using equations \eqref{eq:funksjoner} and \eqref{eq:branch} in appendix \ref{sec:app_contour}. If one
interprets $\mathrm{Im} \, \Gamma_{\lambda \rho \mu}(\bk,E+\im 0^+)$ as a density of
states, the first term describes a continuum above the smallest
gap. However, the second term describes {\it additional discrete states below the
smallest gap}. These are the Andreev bound states induced by the
reflection from the surface. Note that they appear only when the sign
of the two gaps differ, as was also noted in
Ref. \onlinecite{iniotakis}. The energy $E_{0,\hat{\bk}}$ is the positive solution to the equation
\begin{align}
  \label{eq:E0_ligning}
  b_{-,\bk_\parallel} \left(\chi_{+,\bk} \chi_{-,\bk} - E_{0,\hat{\bk}}^2\right) +
  b_{+,\bk_\parallel} \sqrt{|\chi_{+,\bk}|^2 -
    E_{0,\hat{\bk}}^2}\sqrt{|\chi_{-,\bk}|^2 - E_{0,\hat{\bk}}^2} = 0
  \,  
\end{align}\end{widetext}
and is measured relative to the Fermi level. Thus, we get a band of
low energy surface bound states in the part of
momentum space where $\chi_{+,\bk}
\chi_{-,\bk} < 0$. Equation \eqref{eq:E0_ligning} corresponds to equation (14) of Ref. \onlinecite{iniotakis}, but
here we have made no assumption of the particular form of $\bB{\bk}$. 

Let us also comment on what happens in the limit of a singlet
superconductor. From equation \eqref{eq:singlet_triplet}, we see that
this limit corresponds to $\chi_{+,\bk} = \chi_{-,\bk}$. In that case,
there are obviously no Andreev bound states and $\Gamma_{\lambda \rho
  \mu}(\bk,\im \omega_n) = g_{+}(\bk,\im \omega_n)/4 = g_{-}(\bk,\im
\omega_n)/4$. The current $I_\text{qp}$ then equals the
result obtained using bulk Green's function.\cite{werthamer,mahan}

Whereas the limit $\chi_{+,\bk} = \chi_{-,\bk}$ corresponds to a
singlet superconductor, setting the gaps to zero corresponds to a
normal metal. The current will in those cases not depend on the nature
of the  
spin-orbit coupling vector $\bB{\bk}$. The reader may wonder why there are
no remnants of the spin-orbit coupling in these limits. This is a
consequence of the approximation of equal densities of states for the
two bands.

\subsection{\label{sec:Josephson} The Josephson current}
  
The two-particle current is found by inserting the Green's function
\eqref{eq:F_half_space} in equation \eqref{eq:Psi_calc}, giving
\begin{align}
  \label{eq:Psi_total}
  \Psi(\im \omega_\nu) & = \frac{\pi^2}{4} \int^\prime \du \bkhat \int^\prime \du
  \bphat \, |T_{\bk_\text{F},\bp_\text{F}}|^2
  N^\A(\bk_\text{F}) N^\B(\bp_\text{F}) \nonumber \\
  & \times \Bigg\{A_1^{\lambda \gamma} (\bk_\text{F},\bp_\text{F})
    \tilde{S}_1^{\lambda \gamma} (\bk_\text{F},\bp_\text{F},\im \omega_\nu)  \\
 & \quad - \frac{1}{2} A_2^{\lambda \gamma \eta \nu}(\bk_\text{F},\bp_\text{F})
   \tilde{S}_2^{\lambda \gamma \eta \nu}
  (\bk_\text{F},\bp_\text{F},\im \omega_\nu) \nonumber \\
 & \quad - \frac{1}{2} A_3^{\lambda \rho \mu \gamma}
  (\bk_\text{F},\bp_\text{F}) \tilde{S}_3^{\lambda \rho \mu \gamma}
  (\bk_\text{F},\bp_\text{F},\im \omega_\nu) \nonumber \\
 & \quad + \frac{1}{4} A_4^{\lambda \rho \mu \gamma \eta \nu} (\bk_\text{F},\bp_\text{F})
  \tilde{S}_4^{\lambda \rho \mu \gamma \eta \nu}
  (\bk_\text{F},\bp_\text{F},\im \omega_\nu) \Bigg\} \, . \nonumber
\end{align}
As before, the integrals are restricted to positive
$\hat{k}_x,\hat{p}_x$ and repeated greek indices are summed over.
The $\tilde{S}_i$'s are defined as
\begin{align}
\label{eq:Stilde_def}
  & \tilde{S}_1^{\lambda \gamma}(\bk,\bp,\im \omega_\nu) = \frac{1}{\beta} \sum_{\omega_n}
  f^{\ast \,\A}_\lambda(\bk,\im \omega_n - \im \omega_\nu) f^\B_\gamma(\bp,\im
  \omega_n) \, , \nonumber \\
   & \tilde{S}_2^{\lambda \gamma \eta \nu}
  (\bk,\bp,\im \omega_\nu)= \frac{1}{\beta} \sum_{\omega_n}
  f^{\ast \,\A}_\lambda(\bk,\im \omega_n - \im \omega_\nu) \Lambda^\B_{\gamma \eta
    \nu} (\bp,\im \omega_n) \, , \nonumber \\
  & \tilde{S}_3^{\lambda \rho \mu \gamma} (\bk,\bp,\im \omega_\nu) =
  \frac{1}{\beta} \sum_{\omega_n} \Lambda^{\ast \,\A}_{\lambda \rho \mu}(\bk,\im
  \omega_n - \im \omega_\nu) f^\B_\gamma(\bp,\im \omega_n) \, , \nonumber \\
  & \tilde{S}_4^{\lambda \rho \mu \gamma \eta \nu}
  (\bk,\bp,\im \omega_\nu) \\
 & \qquad \qquad = \frac{1}{\beta} \sum_{\omega_n} \Lambda^{\ast \,\A}_{\lambda \rho \mu}(\bk,\im
  \omega_n - \im \omega_\nu) \Lambda^\B_{\gamma \eta
    \nu} (\bp,\im \omega_n). \nonumber 
\end{align} 
The function $f_\lambda(\bk,\im \omega_n)$ is defined in \eqref{eq:quasi} and $\Lambda_{\lambda \rho \mu}(\bk,\im
  \omega_n)$ is 
\begin{equation}
    \label{eq:Lambda_def}
    \Lambda_{\lambda \rho \mu}(\bk,\im
  \omega_n) = \frac{g_\lambda(g^\ast_\rho f_\mu - f_\rho g^\ast_\mu) +
    f_\lambda(f^\ast_\rho f_\mu + g_\rho g^\ast_\mu)}{K(\bk,\im
    \omega_n)} \, .
\end{equation}
As in the previous subsection, we assume that the gaps are
phase-locked, {\it i.e.} $\chi^\A_{\pm, \bk}
\rightarrow \chi^\A_{\pm,\bk} \ex{\im \vartheta^\A}$ and treat
$\chi^{\A(\B)}_{\pm,\bk(\bp)}$ as real. 

In the limit of a singlet superconductor, $\Lambda_{\lambda \rho
  \mu}(\bk,\im \omega_n) = f_{+}(\bk,\im \omega_n)/4 = f_{-}(\bk,\im
\omega_n)/4$. The Josephson current reduces to the result found using
bulk Green's functions.\cite{ambegaokar}

It should be noted that \eqref{eq:Psi_total} is a
tunneling limit expression. Thus, it may not capture all the unusual
phenomena that arise when Andreev bound states contribute to Josephson currents.\cite{lofwander}

\section{\label{sec:numerical_results} Results}

In this section, we consider a junction consisting of two equal
superconductors and present numerical results on the quasiparticle and
Josephson current. We choose to study the Rashba interaction
\begin{equation}
  \label{eq:B_Rashba}
  \bB{\bk} = \alpha \left(\hat{\bo{n}} \times \bk \right),
\end{equation}
both because of its simplicity and its relevance to real
materials like CePt$_3$Si (Ref. \onlinecite{frigeri}) and
Cd$_2$Re$_2$O$_7$ (Ref. \onlinecite{sergienko}). The vector $\hat{\bo{n}}$
represents the direction of broken inversion symmetry of the crystal. 

We restrict ourselves to junctions where $\hat{\bo{n}}^\A$ and $\hat{\bo{n}}^\B$
are perpendicular to the tunneling direction, {\it i.e.}
$\hat{\bo{n}}^{\A(\B)} \cdot \hat{\bo{x}} = 0$. The angle $\zeta$ is defined by
\begin{equation}
  \label{eq:zeta_def}
   \cos \zeta \equiv  \hat{\bo{n}}^\A \cdot
\hat{\bo{n}}^\B \, .
\end{equation}
Of course, from an experimental point of
view, only discrete values of the angle $\zeta$
may be realizable. 

The variation of the current with $\zeta$ is a result of the facts that $\hat{\bo{n}}$ determines the spin structure of the
spin-orbit split bands and that spin is conserved in the tunneling
process. It should be noted that replacing one of the superconductors
by a ferromagnet with magnetization $\bo{M}^\B$ and varying $\hat{\bo{n}}^\A
\cdot \bo{M}^\B$ would not necessarily give similar
conductance variations.\footnote{In the case of a ferromagnet with magnetization $\bo{M}^\B$, combinations like
  $\bB{\bk_\text{F}}^\A \cdot \bo{M}^\B + \bB{\bkbar_\text{F}}^\A \cdot \bo{M}^\B
  = 2 \bb^\A \cdot \bo{M}^\B$ would not contribute when summing over all
  $\bk_\parallel$ and using $\bo{b}_{-\bk_\parallel}^\A = -\bb^\A$. However,
  introducing some spatial anisotropy in the densities of states
  could change this. See also Ref. \onlinecite{molenkamp}, where a
  junction between a 2DEG with Rashba spin-orbit coupling and a
  ferromagnet were considered.}

\subsection{\label{sec:num_res_QP} The quasiparticle current}

We now present numerical results on the quasiparticle current $I_\text{qp}$
given by equation \eqref{eq:Phi_total}. In addition to the choice of
Rashba spin-orbit coupling, we also need the angular dependence of the
gaps $\chi_{\pm,\bk}$. As before, we write the phase explicitly, such
that $\chi_{+,\bk}$ and $\chi_{-,\bk}$ are real.

We consider the same gaps as in
Refs. \onlinecite{hayashi} and \onlinecite{iniotakis}, given by $\eta_{\bk,\mathrm{S}}
= \Psi$ and $\eta_{\bk,\mathrm{T}}
= \Delta |\hat{\bo{n}} \times \hat{\bk}|$. The singlet and triplet
components, $\eta_{\bk,\mathrm{S}}$ and $\eta_{\bk,\mathrm{T}}$, are
defined in equations \eqref{eq:gapOrig} and \eqref{eq:singlet_triplet}. $\Psi$ and $\Delta$ are treated as constants for
simplicity. We also assume that $\Psi \geq 0$ and $\Delta \geq 0$ without loss of generality. The gaps in the spin-orbit split bands are then
\begin{equation}
  \label{eq:gap_eksempel}
  \chi_{\pm,\bk} = \Psi \pm \Delta |\hat{\bo{n}} \times \hat{\bk}|.
\end{equation}
Let us define $q = \Psi/\Delta$. Whereas $\chi_{+,\bk}$ is fully gapped if $q > 0$, the gap
$\chi_{-,\bk}$ contains line nodes if $0 < q < 1$. See
Ref. \onlinecite{hayashi} for details. At this point, we should mention
that other explanations of line nodes in CePt$_3$Si have been put forward.\cite{samokhin_symm_nodes,fujimoto_AF_nodes}

It is also for $q < 1$ that we may expect Andreev bound states at the
surface, since $\chi_{+,\bk} \chi_{-,\bk} < 0$ on a part of the Fermi
surface in that case. However, one should note that formation of Andreev bound
states does not depend on the presence of gap nodes. Isotropic $\chi_{\pm,\bk}$ with
different signs will also result in subgap surface bound states.

\begin{figure}[htbp]
  \centering
  \scalebox{0.22}{
    \hspace{-1cm}
   \includegraphics{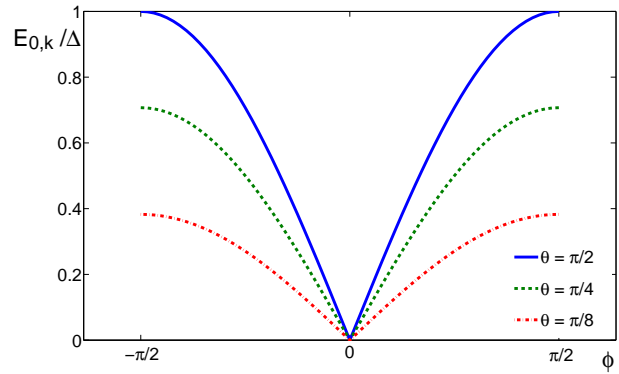}
  }
  \caption{(Color online) Energy spectrum for the bound states
    $E_{0,\hat{\bk}}$ in the case $q = 0$. The dependence on the asimuthal
    angle $\phi$ is shown for three different polar angles $\theta$.}
\label{fig:E0_q0}
\end{figure}

For simplicity, we assume a spherical Fermi surface and let the density
of states be constant over the Fermi surface, $N(\bk_\text{F}) =
N$. Let us introduce spherical coordinates by $\hat{\bk} =  (\cos \phi \sin \theta, \sin \phi
\sin \theta, \cos \theta)$. As mentioned before, the $x$-axis is
perpendicular to the junction. In addition, we let $\hat{\bo{n}}^{\A}$
and $\hat{\bo{n}}^{\B}$ point along or opposite to the
$\hat{\bo{z}}$-direction. The gaps are then given by $\chi_{\pm,\bk}/\Delta = q \pm
\sin \theta$. For $q < 1$, Andreev bound states are formed for momenta with
$\arcsin(q) < \theta < \pi - \arcsin(q)$. As mentioned in section
\ref{sec:QP}, these surface bound states form below the smallest gap,
{\it i.e.} below $|\chi_{-,\bk}|/\Delta = |q - \sin \theta|$. Figure
\ref{fig:E0_q0} shows the spectrum of Andreev bound states $E_{0,\hat{\bk}}$
in the case $q=0$. The dependence on the asimuthal angle $\phi$ is shown for three
different polar angles $\theta$. We see that $E_{0,\hat{\bk}}
\rightarrow 0$ as $\phi \rightarrow 0$, which corresponds to $k_y=0$ and thus $|\bb| = 0$. The maximal
value of $E_{0,\hat{\bk}}$ is given by $|q-\sin \theta|$.

The tunneling matrix element, defined in equations
\eqref{eq:tunn_Ham} and \eqref{eq:T_kp_egenskaper}, will typically
favour momenta with a large component in the tunneling direction. Also, in the case of a
smooth barrier, the parallel momentum is conserved in the tunneling
process. Thus, we assume that
\begin{equation}
  \label{eq:tunn_matrise_eksempel}
  |T_{\bk_\text{F},\bp_\text{F}}|^2 = t \ \hat{k}_x \, \hat{p}_x \, \delta(\hat{\bk}_\parallel - \hat{\bp}_\parallel)
\end{equation}
will capture the qualitative features of the tunneling matrix
element, where $t$ is a constant.\footnote{The size of $t$ could of
  course depend on $\zeta$, at least if different junctions are used
  for different $\zeta$. However, the qualitive features of the
  current-voltage diagram would be unaffected by that.}  

One may show that the variation of the current with the angle $\zeta$
disappears when only perpendicular momenta contribute. In other
words, the effect is dependent on a finite tunneling cone, where also
nonzero parallel momenta contribute to the current.

Tunneling spectroscopy on superconductors are interesting at low
temperatures. At higher temperatures, sharp features giving
information on pairing states may be smeared out. Thus, we investigate
the limit of zero temperature here. However, for $q < 1$, the current
at low voltage is dominated by resonant transport between Andreev bound
states. This is contained in the sum $S_4^{\lambda \rho \mu \gamma \eta \nu}
  (\bk_\text{F},\bp_\text{F},\im \omega_\nu)$, where a product of two delta
  distributions enters. At zero temperature, this leads to a discontinuity at $V=0$,
where the current jumps from zero to a finite value. The discontinuity disappears for nonzero temperatures and a
sharp zero bias conductance peak appears. To get realistic
current-voltage diagrams, we therefore retain a small temperature
($T/\Delta = 0.015$) in this particular term, such that this
discontinuity at zero voltage is smeared out. Such a small temperature will
have no significant effect on the other terms.

The current-voltage diagrams for several $q$ are now presented, where we
have defined $i_\text{qp} \equiv - I_\text{qp}/(2e \pi^2 t^2
N^2)$. We consider the cases of $\zeta = 0$ and $\zeta = \pi$, {\it
  i.e.} equal and opposite directions of broken inversion symmetry. In addition, we present the differential conductance $G(eV) \equiv \du
i_\text{qp} / \du (eV)$, which may be directly accessable in
experiments. The latter has been obtained through a Savitzky-Golay smoothing
filter\cite{press} to remove noise from the numerical integration.

\begin{figure}[htbp]
   \centering
   \scalebox{0.22}{
     \hspace{-1cm}
    \includegraphics{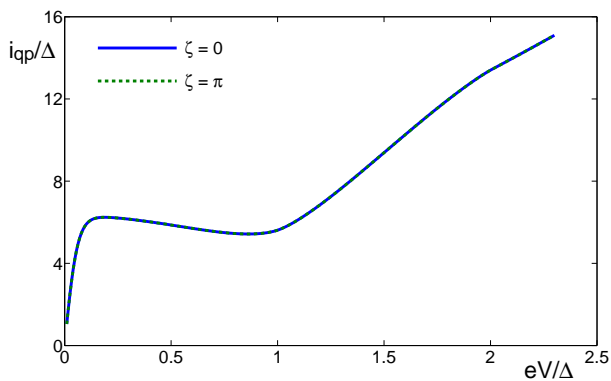}
   }
   \caption{(Color online) Current-voltage diagram in a Josephson
     junction when $q = \Psi/\Delta =
     0$. Transport between Andreev bound states dominates for small
     voltages. There is no dependence on $\zeta$ in this pure triplet case.}
 \label{fig:q0_current}
 \end{figure}

\begin{figure}[htbp]
   \centering
   \scalebox{0.22}{
     \hspace{-1cm}
    \includegraphics{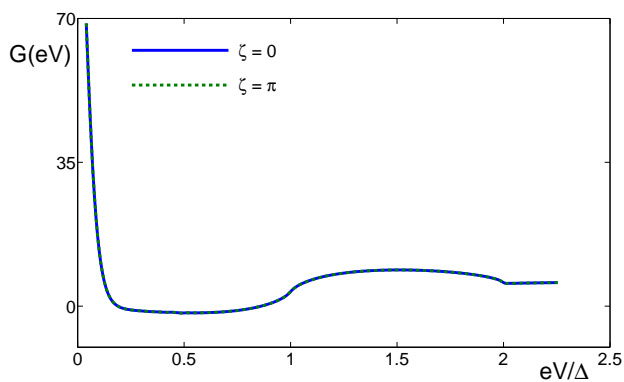}
   }
   \caption{(Color online) Differential conductance as a function of
     voltage when $q=0$. For small voltages, the zero bias conductance peak followed by
     negative differential resistance is due to transport between
     Andreev bound states on each side. There is no dependence on
     $\zeta$ in this pure triplet case.}
 \label{fig:q0_conductance}
 \end{figure}

We start by considering the $q = 0$ case, which corresponds to a pure spin
triplet state. The gaps $\chi_{+,\bk}$ and $\chi_{-,\bk}$ are then of
opposite signs on the entire Fermi surface except at
$\hat{\bk}_\text{F} = \pm \hat{\bo{n}}$, where they have point
nodes. Figure \ref{fig:q0_current} shows the current-voltage diagram
when $q=0$. The differential conductance is
presented in figure \ref{fig:q0_conductance}. The large current at small voltages is due to transport between
Andreev bound states on each side. This gives rise to a zero bias
conductance peak followed by negative differential
conductance. Similar phenomena appear in some $d$-wave
junctions.\cite{lofwander} We observe that there is no difference
between the cases $\zeta = 0$ and $\zeta = \pi$ in the pure triplet case. As stated
before, this is also the case for the pure singlet case, $q
\rightarrow \infty$.\footnote{This statement depends on the two
  densities of states in the nonsuperconducting phase being
  equal. However, if
  they are not, this should give observable effects also above $T_\text{c}$.} However, we will see that this changes for finite $q$, when the gap is a
mixture of singlet and triplet. 

\begin{figure}[htbp]
  \centering
  \scalebox{0.22}{
    \hspace{-1cm}
   \includegraphics{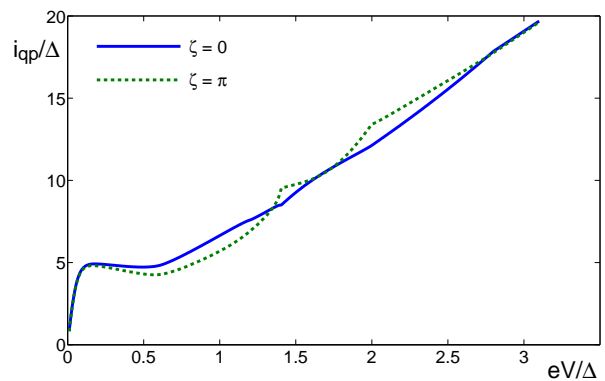}
  }
  \caption{(Color online) Current-voltage diagram when $q = 0.4$. Transport between Andreev bound states dominates for small
     voltages.}
\label{fig:q04_current}
\end{figure}

\begin{figure}[htbp]
  \centering
  \scalebox{0.22}{
    \hspace{-1cm}
   \includegraphics{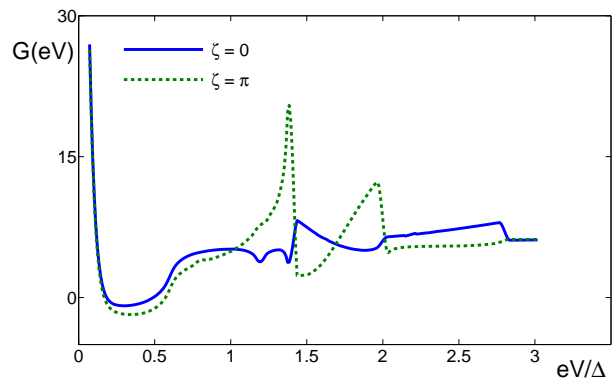}
  }
  \caption{(Color online) Differential conductance in the case $q =
    0.4$. Transport between Andreev bound states dominates for small
     voltages.}
\label{fig:q04_conductance}
\end{figure}

Figure \ref{fig:q04_current} shows the current-voltage diagram in the
case $q = 0.4$. In this case, $\chi_{+,\bk}$ is fully gapped
(although anisotropic) whereas the gap $\chi_{-,\bk}$ has got line
nodes at $\theta \approx 23.6^\circ$ and $\theta \approx
66.4^\circ$. Andreev bound states exist between these angles. 
Observe that the cases $\zeta = 0$ and $\zeta =
\pi$ differs. This becomes clearer when studying the differential
conductance in figure \ref{fig:q04_conductance}. We do not attempt to
explain every feature in this figure, as this depends on the
particular pairing state chosen. In addition, some of these features
might also be smeared out in experimental results. However, the important thing
to notice, which might be observable, is the qualitative difference of
a junction with equal $\hat{\bo{n}}$-vectors ($\zeta = 0$) and one
with opposite $\hat{\bo{n}}$-vectors ($\zeta = \pi$).

\begin{figure}[htbp]
  \centering
  \scalebox{0.22}{
    \hspace{-1cm}
   \includegraphics{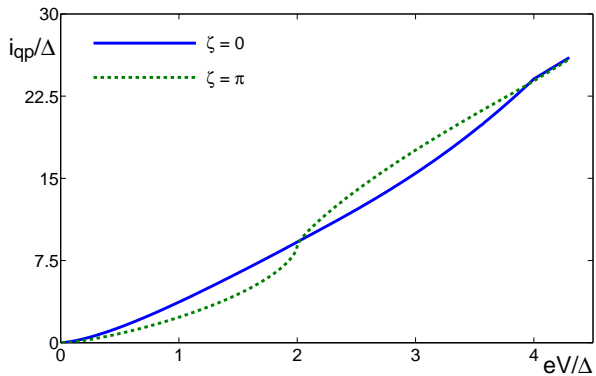}
  }
  \caption{(Color online) Current-voltage diagram when $q = 1$. At
    this point, there are no Andreev bound states.}
\label{fig:q1_current}
\end{figure}

\begin{figure}[htbp]
  \centering
  \scalebox{0.22}{
    \hspace{-1cm}
   \includegraphics{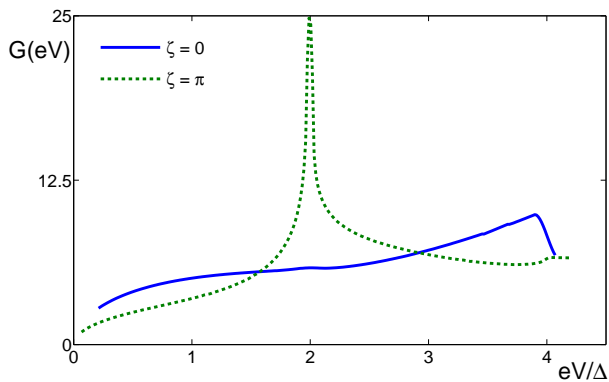}
  }
  \caption{(Color online) Differential conductance when $q = 1$. Note
    the qualitative difference in the two cases for $eV/\Delta < 4$.}
\label{fig:q1_conductance}
\end{figure}

The next current-voltage diagram, presented in figure \ref{fig:q1_current}, is for $q=1$. Then, the line nodes have moved to
the equator ($\theta = \pi/2$) and will disappear for $q > 1$. Now, there
is no part of the Fermi surface where $\chi_{+,\bk}\chi_{-,\bk} < 0$,
such that there are no Andreev bound states. In the differential
conductance in figure \ref{fig:q1_conductance}, there is a clear difference between
$\zeta =0$ and $\zeta = \pi$. See Ref. \onlinecite{borkje} for a
simplified discussion of why this occurs.

\begin{figure}[htbp]
  \centering
  \scalebox{0.22}{
    \hspace{-1cm}
   \includegraphics{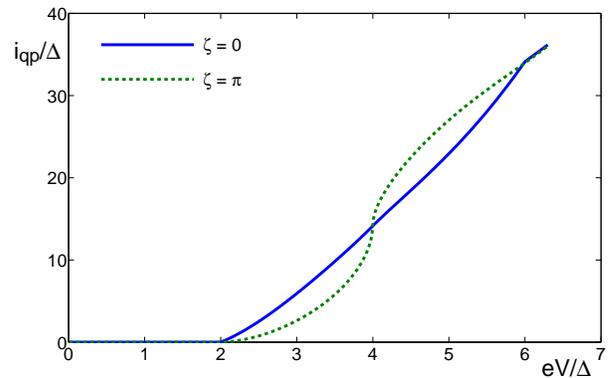}
  }
  \caption{(Color online) Current-voltage diagram when $q = 2$. At
    this point, both bands are fully gapped.}
\label{fig:q2_current}
\end{figure}

\begin{figure}[htbp]
  \centering
  \scalebox{0.22}{
    \hspace{-1cm}
   \includegraphics{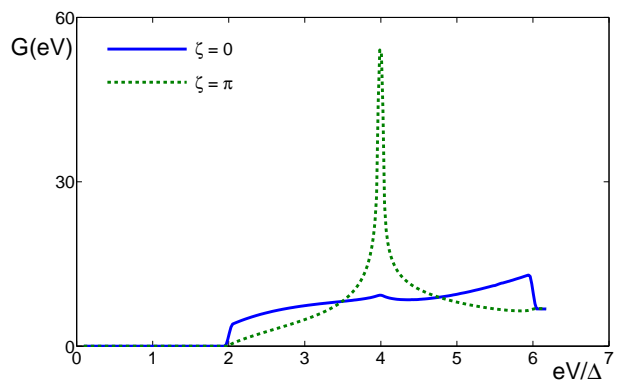}
  }
  \caption{(Color online) Differential conductance when $q = 2$. The
    two cases of $\zeta = 0$ and $\zeta = \pi$ differ significantly when $2 < eV/\Delta <
  6$.}
\label{fig:q2_conductance}
\end{figure}

Finally, we examine the scenario where the singlet to triplet ratio is
$q=2$. At this value, both $\chi_{+,\bk}$ and $\chi_{-,\bk}$ are fully
gapped and of the same sign. The current-voltage diagram is given in figure
\ref{fig:q2_current} and the differential conductance in
\ref{fig:q2_conductance}. Above $eV/\Delta = 2$, the behaviour is
similar to the $q=1$ case. 

In the cases $q=1$ and $q=2$, we observe that the graphs differ in the
region $2(q-1) < eV/\Delta < 2(q+1)$. This will also be the case for
higher values of $q$, but the width of this region ($2(q+1)-2(q-1)=4$) becomes small
relative to the voltages at which the graphs differ ($eV/\Delta
\approx 2q$). In the limit $q \rightarrow \infty$, we are left with
the singlet result, with a single step in the current for both $\zeta = 0$ and $\zeta = \pi$.

\subsection{\label{sec:num_res_Joseph} The Josephson current}

We now move on to the Josephson current, given by equation
\eqref{eq:Psi_total}. This has not been investigated in as much detail as
the quasiparticle current. In this section, we only suggest that
the critical Josephson junction at zero voltage may depend on the
angle $\zeta$ between the axes of broken inversion symmetry of the crystal. We make
no attempt to give any quantitative estimates here, since this depends
not only on the particular pairing state of the material in question, but also on several other issues,
like the details of the tunneling matrix elements. Only
experiments can determine whether this effect really occurs and to what
degree. 

The critical or maximal Josephson current at $eV = 0$, $I_\text{J,c}(\zeta)$ is defined as the absolute
value of the Josephson current at phase difference $\vartheta_\B
- \vartheta_\A = \pm \pi/2$. We still use the Rashba spin-orbit
coupling and consider only one pairing
state, given by
\begin{equation}
  \label{eq:gap_Joseph}
  \chi_{+,\bk} = \text{const.},  \  \chi_{-,\bk} = 0.
\end{equation}
This is probably not very realistic, but suffices to illustrate the
effect.\footnote{These particular gaps have the nice property that the energy integrals
  may be performed analytically, given some reasonable
  approximations.} In this case, there are no Andreev bound states.

As mentioned in the previous subsection, the dependence on $\zeta$
disappears when only perpendicular momenta contribute to the
current. This is also the case for the Josephson current. We
illustrate this by introducing a cutoff in the angle integrals,
integrating over $\theta_\text{c} < \theta < \pi - \theta_\text{c}$ and $-\pi/2 +
\phi_\text{c} < \phi < \pi/2 - \phi_\text{c}$. Here, $\theta_\text{c} =
\phi_\text{c} = 0$ corresponds to integration over the entire
semisphere, whereas only perpendicular momenta contributes when $\theta_\text{c} = \phi_\text{c} \rightarrow \pi/2$.

\begin{figure}[htbp]
  \centering
  \scalebox{0.22}{
    \hspace{-1cm}
   \includegraphics{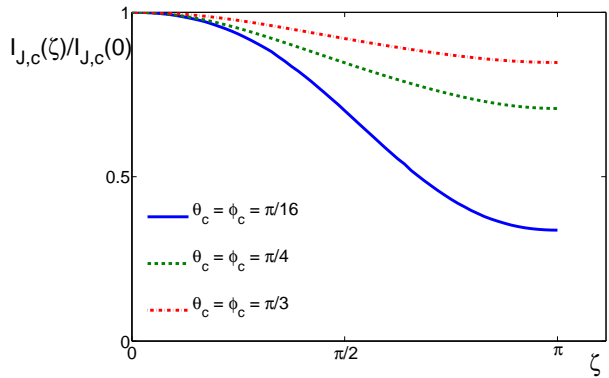}
  }
  \caption{(Color online) The variation of the critical Josephson
    current with $\zeta$ for three different cutoff angles. Note the
    difference in the cases $\zeta = 0$ and $\zeta = \pi$. The
    variation diminishes as the tunneling cone is narrowed.}
\label{fig:Joseph}
\end{figure}

Figure \ref{fig:Joseph} shows the variation of the critical Josephson
current with $\zeta$. Note the difference in current for the cases
$\zeta = 0$ and $\zeta = \pi$. One should also observe that the variation is
reduced when the cutoff angle increases, corresponding to a narrowing
of the
tunneling cone.

Although we have only studied a special scenario, the general message
is that a variation of the critical Josephson current with $\zeta$ may
be expected when the gap is a mixture of singlet and triplet components.

\section{Concluding remarks}

We have investigated both the current-voltage diagram and the critical
Josephson current in
planar tunnel junctions consisting of two superconductors with
antisymmetric spin-orbit coupling. This is relevant for several
recently discovered superconductors, where the spin-orbit coupling is a
consequence of the crystal lacking inversion symmetry. Expressions for the
currents have been derived in the tunneling limit using a general
spin-orbit coupling. 

Numerical results have been presented in the case
of the Rashba spin-orbit coupling. We have investigated the dependence
on the relative angle between
the directions of broken inversion symmetry on each side of the
junction. It has been shown that if the gap
is a mixture of spin singlet and spin triplet parts, qualitative
changes in the differential conductance may be expected when varying this
angle. One may also observe quantitative changes in
the critical Josephson current. This is a result of the fact that
spin is conserved in the tunneling process whereas the spin structure
of the spin-orbit split bands is determined by the direction of broken
inversion symmetry. One should note that broken inversion symmetry on
both sides of the junction is of importance. As stated earlier, similar conductance variations
does not necessarily appear when replacing one of the superconductors with a
ferromagnet and varying its magnetization.

The experimental verification of these phenomena require synthesis of junctions with specific
crystallographic orientation on each side. It is worth mentioning that
Josephson junctions with controllable crystallographic orientation were essential in proving the $d$-wave symmetry of the
order parameter in the high-$T_\text{c}$ cuprates.\cite{tsuei}
Furthermore, the
roughness of the tunnel barrier should be as small as possible. In
addition, the planar tunnel junctions must be thin enough to ensure
that momenta with finite parallel components contribute to the
current. Finally, it should be pointed out that a difference in the
normal phase densities of states of the two bands could give rise to
some of the above mentioned effects even for conventional pairing. However, this should be possible to
detect by measuring the current-voltage characteristics in the normal phase above $T_\text{c}$.

Many approximations and assumptions have been made in order to
produce these results. Thus, the results presented here are expected to be of
qualitative value only. The main message is that experiments on
Josephson junctions of noncentrosymmetric superconductors may provide a direct connection between the
possibly unconventional pairing and the lack of inversion symmetry in
the crystal.

\begin{acknowledgements}
The author would like to thank Yukio Tanaka for a very informative
electronic mail correspondence. Discussions with Asle Sudb{\o}, Thomas Tybell and Eskil K. Dahl have also been
of great value. This work was
supported by the Research Council of Norway, Grant No. 158547/431 (NANOMAT).
\end{acknowledgements}

\appendix
\begin{widetext}
\section{\label{sec:app_Green} Derivation of the half-space Green's function }
First, we note that the momentum space Gor'kov equations in the bulk are
\begin{equation}
  \label{eq:Gorkov_bulk}
  {\cal A} (\bk,\im \omega_n) \Gfour_{\text{b}} (\bk,\im \omega_n)  =
  \unit ,
\end{equation}
where
\begin{equation}
  \label{eq:A_bulk}
  {\cal A} (\bk,\im
  \omega_n)  = \left( \begin{array}{cc}
  \left[\im \omega_n - (\varepsilon_\bk - \mu) \right] \unit - \bB{\bk}
  \cdot \sigma  & -\Delta_\bk \vspace{0.2cm} \\ 
  -\Delta^\dagger_\bk  & \left[\im \omega_n +
  (\varepsilon_\bk - \mu) \right] \unit - \bB{\bk} \cdot \sigma^\ast \end{array} \right)
\end{equation}
 and $\Gfour_{\text{b}}
(\bk,\im \omega_n)$ are matrices in spin times particle-hole space.
The subscript b denotes
bulk. $\omega_n = (2n+1) \pi /\beta$
is a fermion Matsubara frequency. The definition of $\Gfour_{\text{b}}
(\bk,\im \omega_n)$ and the solution of equations
\eqref{eq:Gorkov_bulk} are given in section \ref{sec:Bulk}.

We now want to determine the normal and anomalous Green's function when we
restrict our system to a half-space, {\it i.e} $x < 0$. Contrary to
the bulk case, the Green's function will not be diagonal in momentum
space. We do however assume translational symmetry in the $y$- and
$z$-directions, such that the Green's
function will be diagonal in $\bk_\parallel = k_y \hat{\bo{y}} + k_z
\hat{\bo{z}}$. It is convenient to work in a mixed basis, where we define the Green's functions ${\cal \tilde{G}}_{\sigma \sigma'}(x_1,x_2,\bk_\parallel,\tau) = - \langle
T_\tau \, c_{x_1,\bk_\parallel,
  \sigma}(\tau) \, c^\dagger_{x_2,\bk_\parallel, \sigma'} (0) \rangle$ and ${\cal \tilde{F}}_{\sigma
  \sigma'}(x_1,x_2,\bk_\parallel,\tau) = \langle T_\tau \, c_{x_1,\bk_\parallel,
  \sigma}(\tau) \, c_{x_2,-\bk_\parallel,
  \sigma'} (0) \rangle$. The Gor'kov equations in the continuum limit are 
\begin{equation}
\label{eq:Gorkov1}
  \int_{-\infty}^0 \du x \, {\cal A} (x_1,x,-\im \partial_{x},\bk_\parallel,\im
  \omega_n) \, \tilde{\Gfour} (x,x_2,\bk_\parallel,\im \omega_n)  =
  \delta(x_1-x_2) \, \unit
\end{equation}
in spin $\times$ particle-hole space. We have defined
\begin{equation}
\label{eq:A_matrise}
  {\cal A} (x_1,x,-\im \partial_{x},\bk_\parallel,\im
  \omega_n)  = \left( \begin{array}{cc}
  \left[\im \omega_n \unit - H_\N (x,-\im
  \partial_{x},\bk_\parallel) \right] \delta(x_1-x) & -\Delta (x_1,x,\bk_\parallel) \vspace{0.4cm} \\ 
- \Delta^\dagger(x,x_1,\bk_\parallel) & \left[\im \omega_n \unit + H^\ast_\N(x,-\im
  \partial_{x},-\bk_\parallel) \right] \delta(x_1-x) \end{array} \right) 
\end{equation}
where $\Delta$ and $H_\N$ are 2x2-matrices in spin space. The
4x4 Green's function is
\begin{equation}
\label{eq:4x4Greendef}
 \tilde{\Gfour} (x,x_2,\bk_\parallel,\im \omega_n) = 
  \left( \begin{array}{cc}
   {\cal \tilde{G}}(x,x_2,\bk_\parallel,\im \omega_n) & - {\cal \tilde{F}} (x,x_2,\bk_\parallel, \im \omega_n)
   \vspace{0.3cm} \\
   - {\cal \tilde{F}}^\dagger (x,x_2,\bk_\parallel, \im \omega_n) & - {\cal \tilde{G}}^\text{t} (x_2,x,-\bk_\parallel,-\im \omega_n)
   \end{array} \right) 
\end{equation}\end{widetext}
and should fulfill proper boundary conditions.
The equations are valid for $x_1,x_2 < 0$. The difference from the
full space Gor'kov equations is the restriction $x <
0$ in the integral. The bulk version of equation \eqref{eq:Gorkov1} reduces to \eqref{eq:Gorkov_bulk}.

The pair potential $\Delta(x_1,x,\bk_\parallel)$  in equation \eqref{eq:Gorkov1} should
be determined self-consistently. It is well known that it may differ
from its bulk value near surfaces.\cite{ambegaokar2,buchholtz2} However, we will now apply the usual
approximation\cite{kashiwaya} of replacing the pair potential
by its bulk value. Although this is a crude approximation, it is
expected to give qualitatively correct
results.\cite{kashiwaya,iniotakis}

One way of deriving the half-space Green's function is to consider an
infinite system and then introduce a wall of infinitely strong
nonmagnetic impurities in
order to confine the electrons to one side of the
system.\cite{bobkov} The wall of impurities must ensure that
there is no transport (``hopping'') across the wall and no interaction between the two sides. Since we
use a continuum model, a single plane of impurities at $x=0$ will
provide an impenetrable surface. It will however not prevent
interaction between the two sides due to the possibly nonlocal nature of the
pair potential. Nevertheless, this interaction with ``ghosts''
  on the other side of the impurity wall is tantamount to
  approximating $\Delta(x_1,x,\bk_\parallel)$ by its bulk value. In other words, we construct
  an auxiliary system for $x > 0$ such that a particle in $x_1$
  ``feels'' the pair potential $\Delta_\text{b}(x_1,x,\bk_\parallel)$ from all $x$ as it
  would in the bulk. Thus, in the approximation stated above, we may extend the
$x$-integral in equation \eqref{eq:Gorkov1} to also include positive $x$
and use the bulk pair potential $\Delta_\text{b}(x_1,x,\bk_\parallel)$. However, we must
demand that the boundary condition $\tilde{\Gfour} (x_1,x_2,\bk_\parallel,\im \omega_n) = 0$
for $x_1 = 0$ or $x_2 = 0$ is fulfilled, due to the infinitely strong impurities at $x=0$. 

Having made the above mentioned
approximation, it is easy to show that the {\it ansatz} \begin{widetext}
\begin{equation}
  \label{eq:ansatz}
  \tilde{\Gfour} (x_1,x_2,\bk_\parallel,\im \omega_n) = \tilde{\Gfour}_\text{b}
  (x_1,x_2,\bk_\parallel,\im \omega_n) - \tilde{\Gfour}_\text{b}
  (x_1,0,\bk_\parallel,\im \omega_n) \tilde{\Gfour}^{-1}_\text{b}
  (0,0,\bk_\parallel,\im \omega_n)  \tilde{\Gfour}_\text{b}
  (0,x_2,\bk_\parallel,\im \omega_n)
\end{equation}
satisfies the boundary conditions and the Gor'kov equations. Thus, we have expressed the half-space Green's
function in terms of bulk Green's functions.

Since we desire a description of the system in terms of plane wave states, we
are interested in the Fourier representation of the Green's function
\eqref{eq:4x4Greendef},
\begin{equation}
\tilde{\Gfour} (x_1,x_2,\bk_\parallel,\im \omega_n) =  \int_{-\infty}^\infty
\du k_{1,x} \int_{-\infty}^\infty
\du k_{2,x} \, \Gfour (\bk_{1},\bk_{2},\im \omega_n)
\ex{-\im k_{1,x} x_1 + \im k_{2,x} x_2}.
\end{equation}
Using the Fourier representation of the bulk Green's
function, $\tilde{\Gfour}_\text{b} (x_1,x_2,\bk_\parallel,\im \omega_n) = \int_{-\infty}^\infty
\du k_{x}  \Gfour_\text{b} (\bk,\im \omega_n) \ex{-\im k_{x} (x_1 -
  x_2)}$, we arrive at 
\begin{equation}
\Gfour (\bk_1,\bk_2,\im \omega_n) = \left[\Gfour_\text{b} (\bk_1,\im \omega_n) \delta(k_{1,x} - k_{2,x}) - \Gfour_\text{b} (\bk_1,\im \omega_n) \tilde{\Gfour}^{-1}_\text{b}
  (0,0,\bk_\parallel,\im \omega_n)  \Gfour_\text{b}
(\bk_2,\im \omega_n) \right] \, \delta(\bk_{1,\parallel} - \bk_{2,\parallel}).
\end{equation}
We see that the half-space Green's function differs from the bulk function
by the second term, which is nondiagonal in the perpendicular
components of the momenta.

\section{\label{sec:app_Pauli} Trace calculations}

In section \ref{sec:halfspace}, we defined $\tilde{\sigma}^{\lambda \rho
    \mu}_{\bk_1,\bk_2} = \sigma^\lambda_{\hatbB{\bk_1}}
  \sigma^\rho_{\bb} \sigma^\mu_{\hatbB{\bk_2}} \equiv \beta^{\lambda \rho
    \mu}_{\bk_1,\bk_2} \unit + \bo{\alpha}^{\lambda \rho
    \mu}_{\bk_1,\bk_2} \cdot \bo{\sigma}$, where
  $\sigma^\lambda_{\hatbB{\bk}} = \unit + \lambda \hatbB{\bk} \cdot
  \bo{\sigma}$.
Using the algebra of Pauli matrices, one arrives at 
\begin{align}
  \label{eq:beta_alpha_def}
  \beta^{\lambda \rho
    \mu}_{\bk_1,\bk_2} & = 1 + \lambda \rho (\hatbB{\bk_1}^\A \cdot \bb^\A)
  + \rho \mu (\bb^\A \cdot \hatbB{\bk_2}^\A)
  + \lambda \mu (\hatbB{\bk_1}^\A \cdot \hatbB{\bk_2}^\A)  + \im \lambda
  \rho \mu \hatbB{\bk_1}^\A \cdot (\bb^\A \times \hatbB{\bk_2}^\A) \, ,
  \nonumber \\
  \bo{\alpha}^{\lambda \rho
    \mu}_{\bk_1,\bk_2} & = \lambda \hatbB{\bk_1}^\A + \rho \bb^\A + \mu
  \hatbB{\bk_2}^\A + \im \lambda \rho (\hatbB{\bk_1}^\A \times \bb^\A) + \im
  \rho \mu (\bb^\A \times \hatbB{\bk_2}^\A) + \im \lambda \mu
  (\hatbB{\bk_1}^\A \times \hatbB{\bk_2}^\A)  \\  & \ +   \lambda \rho \mu
  \left[(\bb^\A \cdot \hatbB{\bk_2}^\A) \hatbB{\bk_1}^\A + (\hatbB{\bk_1}^\A
    \cdot \bb^\A) \hatbB{\bk_2}^\A - (\hatbB{\bk_1}^\A
    \cdot \hatbB{\bk_2}^\A) \, \bb^\A \right] \, , \nonumber
\end{align}
and similarly for side B, where $\A \rightarrow \B$, $\lambda,\rho,\mu \rightarrow \gamma,
\eta, \nu$ and $\bk \rightarrow \bp$. We now intend to find the functions $A_i(\bk_\text{F},\bp_\text{F})$ defined in
equation \eqref{eq:A_def}, on which both the one-particle current $I_\text{qp}$ and the two-particle
current $I_\text{J}$ depend. First, note that
\begin{align}
  \label{eq:trace1}
 \Tr \, \sigma^{\lambda}_{\hatbB{\bk}^\A}
    \sigma^\gamma_{\hatbB{\bp}^\B}  & = 2 \left[1 + \lambda \gamma
    (\hatbB{\bk}^\A \cdot \hatbB{\bp}^\B)\right] \nonumber \\
   \Tr \,
    \sigma^{\lambda}_{\hatbB{\bk}^\A} \tilde{\sigma}^{\gamma \eta
      \nu}_{\bp_1,\bp_2} & = 2 \left[\beta^{\gamma \eta
      \nu}_{\bp_1,\bp_2} + \lambda (\hatbB{\bk}^\A \cdot \bo{\alpha}^{\gamma \eta
      \nu}_{\bp_1,\bp_2}) \right]  \\
  \Tr \,
    \tilde{\sigma}^{\lambda \rho \mu}_{\bk_1,\bk_2} \sigma^{\gamma}_{\hatbB{\bp}^\B}  & = 2 \left[\beta^{\lambda \rho \mu}_{\bk_1,\bk_2} + \gamma (\bo{\alpha}^{\lambda \rho \mu}_{\bk_1,\bk_2} \cdot  \hatbB{\bp}^\B ) \right] \nonumber \\
  \Tr \, \tilde{\sigma}^{\lambda \rho \mu}_{\bk_1,\bk_2} \tilde{\sigma}^{\gamma \eta
      \nu}_{\bp_1,\bp_2} & = 2 \left[\beta^{\lambda \rho \mu}_{\bk_1,\bk_2} \beta^{\gamma \eta
      \nu}_{\bp_1,\bp_2} + (\bo{\alpha}^{\lambda \rho \mu}_{\bk_1,\bk_2} \cdot \bo{\alpha}^{\gamma \eta
      \nu}_{\bp_1,\bp_2}) \right] \, , \nonumber
\end{align}
obtained by using $\Tr \, \unit = 2$ and $\Tr \, \sigma_i$ = 0. To
simplify the expressions $A_i(\bk_\text{F},\bp_\text{F})$, some
useful relations are $\hatbB{\bk_\text{F}}^\A \cdot \bb^\A =
\hatbB{\bkbar_\text{F}}^\A \cdot \bb^\A = |\bb^{\A}|^2$ and
$\hatbB{\bk_\text{F}}^\A \cdot \hatbB{\bkbar_\text{F}}^\A = 2
|\bb^{\A}|^2 - 1$. In addition, we are only interested in $A_i(\bk_\text{F},\bp_\text{F})$
as appearing in the Fermi surface integrals \eqref{eq:Phi_total} and
\eqref{eq:Psi_total}. This allows further simplifications when using the
symmetries $\chi_{\lambda,\bk}=\chi_{\lambda,\bkbar}$ and $\hatbB{-\bk}
= -\hatbB{\bk}$. Thus, {\it when appearing in the integrals} \eqref{eq:Phi_total} and
\eqref{eq:Psi_total}, we have
\begin{align}
  \label{eq:A_uttrykk}
  A_1^{\lambda \gamma}(\bk_\text{F},\bp_\text{F}) & = 4 \left[1 + \lambda
  \gamma (\hatbB{\bk_\text{F}}^\A \cdot \hatbB{\bp_\text{F}}^\B)
\right] \nonumber \\
  A_2^{\lambda \gamma \eta \nu}(\bk_\text{F},\bp_\text{F}) & = 4 \left[1
  + \gamma \nu + \eta (\gamma + \nu) |\bbp^{\B}|^2 + \lambda \left(\gamma +
  \nu + 2 \gamma \eta \nu |\bbp^{\B}|^2\right) (\hatbB{\bk_\text{F}}^\A \cdot
  \hatbB{\bp_\text{F}}^\B) + \lambda \eta (1 - \gamma \nu)
  \hatbB{\bk_\text{F}}^\A \cdot \bbp^{\B} \right] \nonumber \\
  A_3^{\lambda \rho \mu \gamma}(\bk_\text{F},\bp_\text{F}) & = 4 \left[1
  + \lambda \mu + \rho (\lambda + \mu) |\bb^{\A}|^2 + \gamma \left(\lambda +
  \mu + 2 \lambda \rho \mu |\bb^{\A}|^2\right) (\hatbB{\bk_\text{F}}^\A \cdot
  \hatbB{\bp_\text{F}}^\B) + \gamma \rho (1 - \lambda \mu)
  \bb^{\A} \cdot \hatbB{\bp_\text{F}}^\B   \right] \nonumber \\
  A_4^{\lambda \rho \mu \gamma \eta \nu}(\bk_\text{F},\bp_\text{F}) &
  = 4 \bigg\{ \left[1
  + \lambda \mu + \rho (\lambda + \mu) |\bb^{\A}|^2 \right] \left[1
  + \gamma \nu + \eta (\gamma + \nu) |\bbp^{\B}|^2\right]  \\
  & \quad \ + \left[1
  - \lambda \mu + \left(2 \lambda \mu + \rho(\lambda + \mu)\right) |\bb^{\A}|^2 \right] \left[1
  - \gamma \nu + \left(2 \gamma \nu + \eta (\gamma + \nu)\right)
  |\bbp^{\B}|^2\right] \nonumber \\
 & \quad \  + \left[(\lambda + \mu + 2 \lambda \rho \mu |\bb^{\A}|^2 )(\gamma +
   \nu + 2 \gamma \eta \nu |\bbp^{\B}|^2 ) + \lambda \gamma + \mu \nu
 \right] (\hatbB{\bk_\text{F}}^\A \cdot \hatbB{\bp_\text{F}}^\B) \nonumber \\
 & \quad \ + \left[(\lambda + \mu + 2 \lambda \rho \mu |\bb^{\A}|^2
   )\eta(1-\gamma \nu) + (\lambda + \mu)\eta (1 + \gamma \nu) \right]
 (\hatbB{\bk_\text{F}}^\A \cdot \bbp^{\B}) \nonumber \\
 & \quad \ + \left[\rho(1-\lambda \mu)(\gamma + \nu + 2 \gamma \eta \nu |\bbp^{\B}|^2
   ) + \rho (1 + \lambda \mu)(\gamma + \nu) \right] (\bb^{\A} \cdot
 \hatbB{\bp_\text{F}}^\B)  \nonumber \\
  & \quad \ + \left[\rho(1-\lambda \mu)\eta(1-\gamma \nu) + \rho (1 + \lambda
    \mu) \eta (1 + \gamma \nu) \right] (\bb^{\A} \cdot
 \bbp^\B) \nonumber \\
 & \quad \ + \left(\lambda \nu + \mu \gamma \right) (\hatbB{\bk_\text{F}}^\A
 \cdot \hatbB{\bpbar_\text{F}}^\B) \nonumber \\
 & \quad \ - \left[\frac{1}{4}\rho(\lambda - \mu) \eta (\gamma - \nu) +
   \left(\lambda \mu + \frac{1}{2} \rho (\lambda + \mu)
   \right)\left(\gamma \nu + \frac{1}{2} \eta (\gamma + \nu) \right)
 \right] (\hatbB{\bk_\text{F}}^\A \times \hatbB{\bkbar_\text{F}}^\A)
 \cdot (\hatbB{\bp_\text{F}}^\B \times \hatbB{\bpbar_\text{F}}^\B)
 \bigg\} \nonumber \, .
\end{align}

\section{\label{sec:app_contour} Matsubara sums}

The fermion Matsubara sums in \eqref{eq:S_def} and \eqref{eq:Stilde_def} may
be converted to contour integrals in the complex plane through the
identity
\begin{equation}
  \label{eq:sum_til_integral}
  \frac{1}{\beta} \sum_{\omega_n} A(\im \omega_n - \im \omega_\nu)
  B(\im \omega_n) = -\frac{1}{2 \pi \im} \oint_C \du z \, A(z - \im \omega_\nu)
  B(z) n_\text{F}(z) \, , 
\end{equation}
for general $A(z-\im \omega_\nu)$ and $B(z)$. The contour $C$ must encircle the poles of the Fermi-Dirac function
$n_\text{F}(z) = (1 + \ex{\beta z})^{-1}$. The functions $A(z - \im
\omega_\nu)$ and $B(z)$ appearing in section \ref{sec:current_theory} will have
branch cuts and possibly poles on the lines $\mathrm{Im} \, z = \im \omega_\nu$ and
$\mathrm{Im} \, z = 0$, respectively. This must be taken into account
when deforming the contour. After the deformation has been performed,
we may let $\im \omega_\nu \rightarrow eV + \im 0^+$. 

The functions entering the sums \eqref{eq:S_def} and
\eqref{eq:Stilde_def} are 
\begin{align}
  \label{eq:funksjoner}
  g^\B_\gamma(z) & = - \frac{z}{\sqrt{|\chi^\B_{\gamma,\bp}|^2 - z^2}}
  \nonumber  \\
  f^\B_\gamma(z) & = \frac{\chi^\B_{\gamma,\bp}
  }{\sqrt{|\chi^\B_{\gamma,\bp}|^2 - z^2}} \, \ex{\im
      \vartheta^\B}  \\
  \Gamma^\B_{\gamma \eta \nu}(z) & = - \frac{z
    \Big(\chi^\B_{\gamma,\bp}\chi^\B_{\eta,\bp} +
      \chi^\B_{\eta,\bp} \chi^\B_{\nu,\bp} - \chi^\B_{\gamma,\bp}
      \chi^\B_{\nu,\bp} - z^2 \Big) \sqrt{|\chi^\B_{-d,\bp}|^2 -
      z^2}}{2 \Big(|\chi^\B_{c,\bp}|^2 - z²\Big) \left[b^\B_{-,\bp_\parallel} \Big(\chi^\B_{+,\bp} \chi^\B_{-,\bp} - z^2\Big) +
  b^\B_{+,\bp_\parallel} \sqrt{|\chi^\B_{+,\bp}|^2 -
    z^2}\sqrt{|\chi^\B_{-,\bp}|^2 - z^2} \right]} \nonumber \\
  \Lambda^\B_{\gamma \eta \nu}(z) & =
  \frac{\left[\chi^\B_{\gamma,\bp}\chi^\B_{\eta,\bp}\chi^\B_{\nu,\bp} -
    z^2 \Big(\chi^\B_{\gamma,\bp} + \chi^\B_{\nu,\bp} -
    \chi^\B_{\eta,\bp}\Big) \right] \sqrt{|\chi^\B_{-d,\bp}|^2 -
      z^2}}{2 \Big(|\chi^\B_{c,\bp}|^2 - z²\Big) \left[b^\B_{-,\bp_\parallel} \Big(\chi^\B_{+,\bp} \chi^\B_{-,\bp} - z^2\Big) +
  b^\B_{+,\bp_\parallel} \sqrt{|\chi^\B_{+,\bp}|^2 -
    z^2}\sqrt{|\chi^\B_{-,\bp}|^2 - z^2} \right]} \, \ex{\im
      \vartheta^\B} \, , \nonumber
\end{align}
where we have defined $c = \text{sgn}(\gamma + \eta + \nu)$ and $d =
\gamma \eta \nu$. 

We choose the branch cuts such that
\begin{equation}
  \label{eq:branch}
  \sqrt{|\chi^\B_{\gamma,\bp}|^2 - (E \pm \im 0^+)^2} =
  \sqrt{|\chi^\B_{\gamma,\bp}|^2 - E^2} \ \Theta\Big[|\chi^\B_{\gamma,\bp}| -
  |E|\Big] \mp \, \im \, \text{sgn}(E) \, \sqrt{E^2
    - |\chi^\B_{\gamma,\bp}|^2} \ \Theta\Big[|E| -
  |\chi^\B_{\gamma,\bp}|\Big] \, ,
\end{equation}
and similarly for side A. 

The functions $g^\B_\gamma(\bp,E + \im 0^+)$ and $\Gamma^\B_{\gamma \eta
  \nu}(\bp,E + \im 0^+)$ have the
property $B(E + \im 0^+) = B(E - \im 0^+)^\ast$. In addition, $\mathrm{Im} \, B(E + \im 0^+)$ is an even
function of $E$. Using this, one finds that the imaginary part of the sums in
\eqref{eq:S_def} may be expressed as
\begin{equation}
  \label{eq:S_i_deformert}
  \mathrm{Im} \, S_i(eV + \im 0^+) = - \frac{\mathrm{sgn}(eV)}{\pi} \int_{-\infty}^\infty \du E \ 
  \mathrm{Im} \, \Big[A(E - |eV| + \im 0^+)\Big] \, \mathrm{Im} \,  \Big[B(E + \im
  0^+)\Big] \,
  \Big[n_\text{F}(E-|eV|) - n_\text{F}(E) \Big].
\end{equation}

The functions $f^\B_\gamma(\bp,E + \im 0^+)$ and $\Lambda^\B_{\gamma
  \eta \nu}(\bp,E + \im 0^+)$ have the property $\tilde{B}(E \pm \im 0^+) =
\left[\tilde{B}_\text{R}(E) \pm \im \tilde{B}_\text{I}(E)\right]\ex{\im \vartheta^\B}$,
where the real functions $\tilde{B}_\text{R}(E)$ and
$\tilde{B}_\text{I}(E)$ are even and odd in $E$, respectively. At $eV=0$, this enables us to write the imaginary part
of the sums in
\eqref{eq:Stilde_def} as
\begin{equation}
  \label{eq:Stilde_i_deformert}
  \mathrm{Im} \, \tilde{S}_i(\im 0^+) = \frac{\sin (\vartheta^\B -
    \vartheta^\A)}{\pi} \int_{-\infty}^\infty \du E \,
  \Big[\tilde{A}_\text{R}(E) \tilde{B}_\text{I}(E) + \tilde{A}_\text{I}(E) \tilde{B}_\text{R}(E)
  \Big] \, \Big[1 - 2 n_\text{F}(E) \Big] \, .
\end{equation}



\end{widetext}


\end{document}